\documentclass[11pt]{article}
\usepackage[T1]{fontenc}
\usepackage{amssymb,amsmath,amsfonts,dsfont,amsthm,bm,bbm}
\usepackage{float,rotfloat} 
\usepackage{mathrsfs}
\usepackage{multirow}
\usepackage{enumerate}
\usepackage{enumitem}
\usepackage{mathtools}
\usepackage{hyperref,url}
\usepackage{graphicx}
\usepackage[hang,small,bf]{caption}
\usepackage[top=1in, bottom=1in, left=1.0in, right=1.0 in]{geometry}
\usepackage{longtable}
\usepackage{color,xcolor}
\usepackage[round,sort,comma]{natbib}
\newcounter{bibcount}
\usepackage[title]{appendix}
\usepackage{etoolbox} 
\usepackage[english]{babel}
\usepackage[autostyle, english = american]{csquotes}
\MakeOuterQuote{"}
\patchcmd{\appendices}{\quad}{: }{}{} 
\newtheorem{thm}{Theorem}

\newtheorem{defn}{Definition}
\numberwithin{thm}{section}
\numberwithin{lemma}{section}
\numberwithin{defn}{section}
\numberwithin{equation}{section}
\numberwithin{figure}{section}
\numberwithin{table}{section}
\newtheorem{note}{Note}[section]

\newcommand{\ID}{\mathds{1}}

\definecolor{darkblue}{rgb}{0,0,0.55}

\makeatletter
\patchcmd{\@lbibitem}%
{\item[}%
{\item[\hfill\stepcounter{bibcount}{[\thebibcount]}}%
{}%
{}
\patchcmd{\@lbibitem}%
{\hfil \NAT@anchor {#2}{\NAT@num }]}%
{\hyper@natanchorstart{#2\@extra@b@citeb}\hyper@natanchorend]}
{}%
{}
\renewcommand\NAT@bibsetup[1]{%
	\setlength{\leftmargin}{0.40in}
	\setlength{\itemindent}{-\parindent}%
	\setlength{\itemsep}{\bibsep}%
	\setlength{\parsep}{\z@}}
\makeatother

\hypersetup{,colorlinks=true,allcolors=darkblue}

\allowdisplaybreaks

\begin{document}   
\let\nofiles\relax

\baselineskip 5mm

\thispagestyle{empty}

\begin{center}
{\LARGE Method of Winsorized Moments for Robust Fitting
\\[1ex]
of Truncated and Censored Lognormal Distributions}

\vspace{12mm}

{\large\sc
Chudamani Poudyal\footnote[1]{~{\sc Corresponding Author}: 
Chudamani Poudyal, Ph.D., is a Visiting Assistant Professor 
in the Department of Statistics and Data Science, University 
of Central Florida, Orlando, FL 32816, USA. 
~~ {\em e-mail\/}: ~{\tt Chudamani.Poudyal@ucf.edu}}}

\vspace{1mm}

{\large\em University of Central Florida}

\vspace{4mm}

{\large\sc
Qian Zhao\footnote[2]{~Qian Zhao, Ph.D., ASA, is an Assistant 
Professor in the Department of Mathematics, Robert Morris 
University, Moon Township, PA 15108, USA. ~~ {\em e-mail\/}: 
~{\tt zhao@rmu.edu}}}

\vspace{1mm}

{\large\em Robert Morris University}

\vspace{4mm}

{\large\sc  
Vytaras Brazauskas\footnote[3]{~Vytaras Brazauskas, Ph.D., ASA,
is a Professor in the Department of Mathematical Sciences,  
University of Wisconsin-Milwaukee, Milwaukee, WI 53201, USA. 
~~ {\em e-mail\/}: ~{\tt vytaras@uwm.edu}}} 

\vspace{1mm}

{\large\em University of Wisconsin-Milwaukee}

\vspace{5mm}

\copyright \
Copyright of this Manuscript is held by the Authors! 

\end{center}

\vspace{5mm}

\begin{quote}
{\bf\em Abstract\/}.
When constructing parametric models to predict the cost of future 
claims, several important details have to be taken into account:
($i$) models should be designed to accommodate deductibles, 
policy limits, and coinsurance factors,
($ii$) parameters should be estimated robustly to control the 
influence of outliers on model predictions, and
($iii$) all point predictions should be augmented with estimates
of their uncertainty.
The methodology proposed in this paper provides a framework for 
addressing all these aspects simultaneously.
Using payment-per-payment and payment-per-loss variables, 
we construct the adaptive version of 
{\em method of winsorized moments\/} (MWM) estimators 
for the parameters of truncated and censored lognormal distribution.
Further, the asymptotic distributional properties of this approach
are derived and compared with those of the {\em maximum likelihood 
estimator\/} (MLE) and {\em method of trimmed moments\/} 
(MTM) estimators. 
The latter being a primary competitor to MWM.
Moreover, the theoretical results are validated with extensive 
simulation studies and risk measure sensitivity analysis. Finally, 
practical performance of these methods is illustrated using the 
well-studied data set of 1500 U.S. indemnity losses. 
With this real data set, it is also demonstrated
that the composite models do not provide much 
improvement in the quality of predictive models 
compared to a stand-alone fitted distribution 
specially for truncated and censored sample data.

\vspace{4mm}

\end{quote}

\newpage

\setcounter{page}{1} 

\baselineskip 7mm

\section{Introduction}
\label{sec:Introduction}

Parametric inference for loss models is commonly based on 
{\em maximum likelihood estimators\/} (MLEs), which are 
sensitive to model misspecification and outliers. 
To address these vulnerabilities one can employ robust 
techniques such as {\em method of trimmed moments\/} (MTM) 
and {\em method of winsorized moments\/} (MWM); 
see \cite{MR2497558} and 
\cite{MR3758788,MR3750626}.
MTM and MWM, however, were designed for completely observed 
data, not for insurance payment variables which often are 
affected by deductibles, policy limits, or coinsurance factors. 
MLEs can certainly handle such data transformations, but as will 
be shown later even when data are truncated and censored -- thus 
within finite range -- sensitivity of MLEs to data perturbations 
at the interval endpoints is inescapable. Therefore, the main 
motivation for this work is to develop robust estimators for 
the parameters of lognormal distributions when data are
left-truncated and right-censored (these are primary 
transformations for defining insurance payments).

Applications of the lognormal distribution are diverse and 
include actuarial science, business, economics, and other areas 
\citep[see, e.g.,][and the references cited therein]{MR1987777}.
The model has been effectively used with homogeneous loss data 
\citep{hl79,MR3836650} 
as well as with heterogeneous data. 
In the latter case, it has been shown by 
\cite{MR2188658},
\cite{MR2347211},
\cite{MR3474025},
\cite{tcf21},
\cite{MR3543061},
\cite{MR3836650},
\cite{MR3896968}, 
and 
\cite{MR4118953}
that lognormal models can capture the nature of the data set 
either in the head or in the tail or both parts. Also, using 
approximate Bayesian computation methods and Bayesian Information 
Criterion, \cite{MR4287445} 
demonstrated that lognormal distribution provides the best 
fit to claim sizes when compared to gamma and Weibull models.

Further, sensitivity of MLEs to the underlying modeling assumptions 
has been known for a long time; at least since the early 1960s 
\citep{MR0120720}.
While there exist multiple reasons for model misspecification, in the 
actuarial literature researchers emphasized unobserved heterogeneity 
of claims, multi-modality, and different tail behavior of small and 
large claim amounts
\citep[see][]{MR4340261}.
The important 
conclusion that can be drawn from these observations is that MLE-based 
model estimates are usually flawed, resulting in biased risk predictions 
and/or unreliable assessment of their variability. 
One of the most 
popular proposals to deal with such a problem that emerged in the 
literature is to fit spliced (or mixtures of) loss distributions; 
see
\cite{MR2188658}, 
\cite{MR2347211},
\cite{MR3474025},
\cite{MR3896968},
\cite{MR4118953},
\cite{MR4163087},
\cite{MR4340261}. 
In particular,  \cite{MR2842558}, 
\cite{MR3177097}, and \cite{MR4149559} used the lognormal distribution as one component 
in mixed composite models to fit the 
observed claim severtity data sets.
This approach is intuitively appealing because it provides more 
flexibility to the data modeler as mixtures of distributions are 
certainly more flexible than a single loss distribution. However, 
it still remains to be seen whether such an approach yields stable 
cost predictions (i.e., stable against outliers and other data 
perturbations) because fitting of mixture models to data is done 
using MLEs or equivalent methods.

Furthermore, there are several types of statistics that can be used 
to construct robust estimation procedures. In the actuarial literature,
the first comprehensive paper on robust estimation of lognormal 
distributions used generalized $L$-statistics 
\citep{MR1987777}. 
To redesign such computationally intense methods for insurance 
payments, however, is not simple. Therefore, we will employ less 
general (but very effective) techniques based on $L$-statistics 
\citep{MR0203874}. 
In this line of research, 
the robust $t$-score methodology and 
its variants 
\citep{MR1862941, MR2412617, Fab10, MR2720398}
have been designed for completely observed ground-up heavy tailed insurance data. 
For incomplete loss data, several new proposals 
have been made by 
\cite{MR4263275,MR4192140}, 
where the first paper 
developed MTM estimators of the lognormal 
distribution parameters for payment-per-payment
and payment-per-loss 
data scenarios and the second studied robust estimation via data 
truncation, censoring, and related variants. 
In the current literature, MTM is the most effective approach 
for robust estimation of truncated and censored lognormal models. 
But as earlier theoretical and empirical studies have shown 
\citep[see][]{MR3758788,MR3750626}, for complete data 
and under lognormal distributional assumptions, MWM outperforms 
MTM in terms of asymptotically smaller variance (all else being 
equal). 
Thus, in this paper we will develop MWM estimators for the parameters of 
left-truncated and right-censored lognormal distribution, 
and compare them with the corresponding MLE and MTM estimators.
The asymptotic distributional properties 
of MWM are derived and compared with those of MLE and MTM. 
Moreover, the theoretical results are validated with extensive 
simulation studies and risk measure sensitivity analysis. They
are also illustrated on the well-studied data set of 1500 U.S. 
indemnity losses.

The rest of the paper is organized as follows. 
In Section \ref{sec:PP_DataScenarios1}, we briefly summarize 
two types of insurance benefit payments when the underlying 
distribution is lognormal. 
Section \ref{sec:MLE} provides the existing MLE results for 
the two payment variables along with the estimators' asymptotic
distributional properties.
Section \ref{sec:MLE} also includes MLE composite 
models for those two payment data types. 
Section \ref{sec:MWM} is focused on the development of MWM 
procedures for the location and scale parameters under 
the distributional scenarios of Section \ref{sec:PP_DataScenarios1}. 
Also in this section, comparisons of {\em asymptotic relative 
efficiencies\/} of MWM estimators with respect to MLE as well 
as MTM are presented. 
In Section \ref{sec:SimStudy}, we conduct a simulation study
and risk sensitivity study to complement the theoretical 
results. 
Further, the newly designed methodology is implemented on real 
data and its performance is provided in 
Section \ref{sec:RealDataAnalysis}.
Finally, concluding remarks and future outlook are offered 
in Section \ref{sec:Conclusion}.

\section{Payment Data Scenarios}
\label{sec:PP_DataScenarios1}

Consider a ground-up lognormal loss random variable 
$W \sim LN(w_{0},\theta,\sigma)$ with 
cdf $F_{W}$ and pdf $f_{W}$,
\begin{align}
\label{eqn:LNCDF_PDF}
F_{W}(w) 
& = 
\Phi 
\left( 
\dfrac{\log{(w-w_{0}})-\theta}
{\sigma}
\right)
\quad 
\mbox{and} 
\quad 
f_{W}(w) 
= 
\dfrac{1}{\sigma (w-w_{0})} 
\phi 
\left( 
\dfrac{\log{(w-w_{0}})-\theta}
{\sigma}
\right),
\end{align}
for 
$w > w_{0}$, $\sigma > 0$, and $\theta \in \mathbb{R}$,
where 
$\Phi$ and
$
\phi(x) 
= 
\left(
1/\sqrt{2\pi} \,
\right) 
e^{-x^2/2}, \ 
x \in \mathbb{R}
$
are the cdf and pdf of the standard normal distribution, respectively.
Then with policy deductible $d$, policy limit $u$, 
and coinsurance factor $c$, the two typical insurance 
payments random variables, 
{\em payment-per-payment\/} $(Y_w)$ and 
{\em payment-per-loss\/} $(Z_w)$, 
are defined as \citep[see][p. 126]{MR3890025} 
\begin{align}
Y_{w} 
& : =
c
\left(
\min\big\{W,u \big\}-d 
\right)
\, | \, W > d
= 
\left\{ 
\begin{array}{ll}
\mbox{undefined},
& W \le d; \\
c \left( W-d \right), & d < W < u; \\
c \left( u-d \right), & u \le W. \\
\end{array}
\right.
\label{p1data} \\[5pt]
Z_{w}
& := 
c \left( \min \big\{ W, u \big\} - \min \big\{ W, d \big\} \right) 
= 
\left\{ 
\begin{array}{ll}
0, & W \leq d; \\
c \left( W-d \right), & d < W < u; \\
c \left( u-d \right), & u \leq W. \\
\end{array}
\right.
\label{p2data}
\end{align}
Clearly 
$X := \log{(W-w_{0})}\sim N(\theta,\sigma^2)$ 
with cdf and pdf, respectively, given by:
\begin{align}
\label{eqn:LNCDF2}
F(x) 
= 
\Phi\left(\frac{x-\theta}{\sigma}\right) 
\quad \mbox{and} \quad 
f(x) 
= 
\frac{1}{\sigma}
\phi\left(\frac{x-\theta}{\sigma}\right),
\qquad 
-\infty < x < \infty.
\end{align}
The corresponding quantile function 
$F^{-1} : (0,1) \to \mathbb{R}$ is given by 
$
F^{-1}(v) = \theta + \sigma \Phi^{-1}(v).
$
Consider the following notations from \cite{MR4263275}:
\begin{equation}
\label{eqn:uToTdTotDefn1}
t := \log{(d-w_{0})}, \
T := \log{(u-w_{0})}, \
R := T-t, \ 
\gamma := \frac{t-\theta}{\sigma}, \ \mbox{and} \ 
\xi := \frac{T-\theta}{\sigma}.
\end{equation}
Note that it is possible to have $t < 0$ but $d > 0$.
Then, it follows that
\begin{align}
\label{eqn:defTz}
\theta 
& = 
t-\sigma \gamma
\quad \mbox{and} \quad 
\xi 
= 
\gamma+\frac{R}{\sigma}.
\end{align} 
With these notations, the corresponding normal form of the 
random variables $Y_{w}$ and $Z_{w}$, respectively, 
defined by equations \eqref{p1data} and \eqref{p2data} are 
now, respectively, given by:
\begin{align}
Y
& : =
c\log
{
\left( 
\dfrac{Y_{w}}{c(d-w_{0})}
+ 1
\right)
}
=
c
\left(
\min\big\{ X,T \big\}-t
\right)
\, | \, X > t
=
\left\{ 
\begin{array}{ll}
\mbox{undefined}, 
& X \le t; \\
c \left( X-t \right), & t < X < T; \\
c \left( T-t \right), & T \le X. \\
\end{array}
\right.
\label{p1Ndata} \\[5pt]
Z
& : =
c\log
{
\left( 
\dfrac{Z_{w}}{c(d-w_{0})}
+ 1
\right)
}
=
c
\left(
\min\big\{ X,T \big\}
-
\min\big\{ X,t \big\}
\right)
=
\left\{ 
\begin{array}{ll}
0, 
& X \le t; \\
c \left( X-t \right), & t < X < T; \\
c \left( T-t \right), & T \le X. \\
\end{array}
\right.
\label{p2NZdata}
\end{align}

The pdf's of the random variables $Y$ and $Z$
are, respectively, given by:
\begin{align}
\label{p1pdf}
f_{Y}( y; \, c, t, T ) 
& = 
\left\{
\begin{array}{ll}
\frac{f_{X}(y/c+t)}{c [1-F_{X}(t)] }, & 0 < y < c(T-t); \\[1ex]
\frac{1-F_{X}(T^-)}{1-F_{X}(t)}, & y = c(T-t); \\[0.75ex]
0, & \mbox{elsewhere}. \\
\end{array}
\right. \\[5pt]
f_{Z}( z; \, c, t, T ) 
\label{p2pdf}
& =  
\left\{
\begin{array}{ll}
F_{X}(t), & z = 0; \\[0.25ex]
f_{X}(z/c+t)/c, & 0 < z < c(T-t); \\[0.25ex]
1 - F_{X}(T^-), & z = c(T-t); \\[0.25ex]
0, & \mbox{elsewhere}. \\
\end{array}
\right.
\end{align}

\begin{note}
The constants $t$ and $T$ can be treated as transformed
deductible and policy limit, respectively,
for the normal random variable 
$X \sim N(\theta, \sigma^2)$ with a possibility 
of $t < 0$.
\qed 
\end{note}

\section{MLE}
\label{sec:MLE}

If a truncated (both singly and doubly) normal 
sample data set is available then the MLE procedures  
for such data have been developed by \cite{MR0038041} 
and the method of moments estimators can be found in 
\cite{MR0045361} and \cite{MR0196848}.
The corresponding results for payments 
$Y$ and $Z$ data types have been established 
by \cite{MR4263275}.

For any $0 \le s \le 1$, define 
$\bar{s} = 1-s$. Therefore,  
$
\bar{\Phi}(z)
= 
1 - \Phi(z)
$
be the standard normal survival function at 
$z \in \mathbb{R}$.
Consider
\begin{align}
\label{eqn:truncatedNormalZDef}
\Omega_{1} 
& :=
\frac{\phi(\gamma)}{\bar{\Phi}(\gamma)-\bar{\Phi}(\xi)}
\quad \mbox{and} \quad 
\Omega_{2} 
:= 
\frac{\phi(\xi)}{\bar{\Phi}(\gamma)-\bar{\Phi}(\xi)}.
\end{align}

\begin{note} 
For MLE estimation purposes, the variable $\gamma$, defined 
in (\ref{eqn:uToTdTotDefn1}), can be treated the same way as 
parameter $\theta$. Therefore, the mean $\theta$ is a linear 
function of $\gamma$ given by (\ref{eqn:defTz}).
\qed 
\end{note}

\subsection{Payments {\em Y}}

Let $y_{1},\ldots,y_{n}$ be an {\em i.i.d.\/} sample 
given by pdf $f_{Y}$ with policy limit $T$, 
deductible $t$, and coinsurance factor $c$. 
Define
$
n_{1} 
:=
\sum_{i=1}^{n}\ID\{0<y_{i}<cR\} \ \mbox{and} \
n_{2} 
:= 
\sum_{i=1}^{n}\ID\{y_{i}=cR\}
$.
Then the MLE system of implicit equations to be
solved for 
$
\left( 
\widehat{\gamma}_{\mbox{\tiny y,MLE}},
\widehat{\sigma}_{\mbox{\tiny y,MLE}}
\right) 
$
is given by:
\begin{align}
\label{eqn:PPNormalMLEeqn2}
\left\{
\begin{array}{rrr}
\sigma\left(\Omega_{y,1}-\Omega_{y,2}-\gamma\right)
- c^{-1}\widehat{\mu}_{y,1} 
& = & 0, \\[10pt]
\sigma^{2}\left(1-\gamma(\Omega_{y,1}-\Omega_{y,2}-\gamma)
-\frac{\Omega_{y,2}R}{\sigma}\right) -  c^{-2}\widehat{\mu}_{y,2} 
& = & 0,
\end{array} \right.
\end{align}
where 
$\widehat{\mu}_{y,1}$ and $\widehat{\mu}_{y,2}$ 
are the first and second sample moments, 
$\widehat{\mu}_{y,j} 
:=
n_{1}^{-1}\sum_{i=1}^{n}\ID\{0<y_{i}<cR\}y_{i}^{j}$, $j = 1,2$,
and
\begin{align}
\label{eqn:PPNormalQDef}
\Omega_{y,1} 
& :=
\frac{n}{n_{1}} \frac{\phi(\gamma)}{\bar{\Phi}(\gamma)}
\quad \mbox{and} \quad 
\Omega_{y,2} 
:=
\frac{n_{2}}{n_{1}} \frac{\phi(\xi)}{\bar{\Phi}(\xi)}, 
\end{align}
Further it has been established by \cite{MR4263275}
that 
\begin{equation}
\label{eqn:MLEYDeltaMatrix1}
(\widehat{\gamma}_{\mbox{\tiny y,MLE}},
\widehat{\sigma}_{\mbox{\tiny y,MLE}}) 
\sim 
\mathcal{AN}\left((\gamma,\sigma),
\frac{\Lambda^{-1}}{n\left(r_{1}r_{3}-r_{2}^{2}\right)}
\begin{bmatrix}
-r_{3} & \sigma r_{2} \\[10pt]
\sigma r_{2} & -\sigma^{2}r_{1} 
\end{bmatrix}
\right),
\end{equation}
where 
$\Lambda 
:=
\frac{\bar{\Phi}(\gamma)-\bar{\Phi}(\xi)}
{\bar{\Phi}(\gamma)}$ and
\begin{align}
\label{eqn:PPNormalMLEFishergFunctions}
\left\{
\begin{array}{lll}
r_{1}(\gamma,\xi) 
& := & 
-\left[1+\gamma\Omega_{1}-\xi\Omega_{2}
-
\frac{\phi(\gamma)}{\bar{\Phi}(\gamma)}\Omega_{1}  +
\frac{\phi(\xi)}{\bar{\Phi}(\xi)}\Omega_{2}\right], \\[10pt]
r_{2}(\gamma,\xi) 
& := &
\frac{R\Omega_{2}}{\sigma}
\left[\frac{\phi(\xi)}{\bar{\Phi}(\xi)}
-\xi\right]+\left[\Omega_{1}-\Omega_{2}-\gamma\right], \\[10pt]
r_{3}(\gamma,\xi) 
& := &
\left(\frac{R}{\sigma}\right)^{2}\Omega_{2}
\left(\xi-\frac{\phi(\xi)}{\bar{\Phi}(\xi)}\right)-
\left[2 -\gamma(\Omega_{1}-\Omega_{2}-\gamma)-\frac{\Omega_{2}R}{\sigma}\right]. \\
\end{array} \right.				
\end{align}
Since 
$
(\theta,\sigma)
=
(t-\sigma \gamma, \sigma),
$
then by multivariate delta method
\citep[see, e.g.,][p. 122]{MR595165},
we have
$
(\widehat{\theta}_{\mbox{\tiny y,MLE}},
\widehat{\sigma}_{\mbox{\tiny y,MLE}}) 
\sim 
\mathcal{AN}
\left(
(\theta,\sigma),
\frac{1}{n} \bm{S}_{\mbox{\tiny {y,MLE}}}
\right),
$
where 
\begin{align}
\label{eqn:MLEYDeltaMatrix2}
\bm{S}_{\mbox{\tiny {y,MLE}}}
= 
\frac{\Lambda^{-1}}{\left(r_{1}r_{3}-r_{2}^{2}\right)}
\mathbf{D}
\begin{bmatrix}
-r_{3} & \sigma r_{2} \\[10pt]
\sigma r_{2} & -\sigma^{2}r_{1} 
\end{bmatrix}
\mathbf{D}'
\quad \mbox{and} \quad 
\mathbf{D}
& =
\begin{bmatrix}
-\sigma & -\gamma \\
0 & 1
\end{bmatrix}.
\end{align}

\subsection{Payments {\em Z}}

Consider an observed {\em i.i.d.\/} sample $z_{1},\ldots,z_{n}$
given by pdf $f_{Z}$.
Define
\begin{align}
\label{defn:n0n1n2PPZ}
n_{0} & := \sum_{i=1}^{n}\ID\{z_{i}=0\}, 
& n_{1} & := \sum_{i=1}^{n}\ID\{0<z_{i}<cR\}, 
& n_{2} & := \sum_{i=1}^{n}\ID\{z_{i}=cR\}.
\end{align}
Note that $n=n_{0}+n_{1}+n_{2}$. 
Similarly with (\ref{eqn:truncatedNormalZDef}), define
\begin{align}
\label{eqn:censoredNormalYDef}
\Omega_{z,1} 
& :=
\frac{n_{0}}{n_{1}}\frac{\phi(\gamma)}{{\Phi}(\gamma)}, 
\qquad 
\Omega_{z,2} 
:=
\frac{n_{2}}{n_{1}}{\frac{\phi(\xi)}{\bar{\Phi}(\xi)}}.
\end{align}
Then, the MLE system of equations 
to be solved for $\gamma$ and $\sigma$ becomes
\begin{align}
\label{eqn:censoredNormalMLEeqn2}
\left\{
\begin{array}{rrr}
\sigma\left(\Omega_{z,1}-\Omega_{z,2}-\gamma\right) 
- c^{-1}\widehat{\mu}_{z,1} & = & 0, \\[10pt]
\sigma^{2}\left(1-\gamma(\Omega_{z,1}-\Omega_{z,2}
-\gamma)-\frac{\Omega_{z,2}R}{\sigma}\right)
- c^{-2}\widehat{\mu}_{z,2} & = & 0,
\end{array} \right.				
\end{align}
where 
$\widehat{\mu}_{z,1}$ and
$\widehat{\mu}_{z,2}$ 
are the first and second sample moments, 
$\widehat{\mu}_{z,j} 
:= 
n_{1}^{-1}\sum_{i=1}^{n}\ID\{0<z_{i}<cR\}z_{i}^{j}$, $j=1,2.
$
Further, it has been established by 
\cite{MR4263275} that
\begin{equation}
\label{eqn:MLEZDeltaMatrix1}
(\widehat{\gamma}_{\mbox{\tiny z,MLE}},
\widehat{\sigma}_{\mbox{\tiny z,MLE}}) 
\sim \mathcal{AN}\left((\gamma,\sigma),
\frac{\Lambda^{-1}}{n \bar{\Phi}(\gamma)
\left(\psi_{1}\psi_{3}-\psi_{2}^{2}\right)}
\begin{bmatrix}
-\psi_{3} & \sigma \psi_{2} \\[10pt]
\sigma \psi_{2} & -\sigma^{2}\psi_{1} 
\end{bmatrix}
\right),
\end{equation}
where 
\begin{align}
\label{eqn:censoredNormalMLEFishergFunctions}
\left\{
\begin{array}{lll}
\psi_{1}(\gamma,\xi) 
& := &
-\left[1+\gamma\Omega_{1}-\xi\Omega_{2}
+\frac{\phi(\gamma)}{\Phi(\gamma)}\Omega_{1}  +
\frac{\phi(\xi)}{\bar{\Phi}(\xi)}\Omega_{2}\right], \\[10pt]
\psi_{2}(\gamma,\xi) 
& := &
\frac{R\Omega_{2}}{\sigma}\left[\frac{\phi(\xi)}{\bar{\Phi}(\xi)}-
\xi\right]+\left[\Omega_{1}-\Omega_{2}-\gamma\right], \\[10pt]
\psi_{3}(\gamma,\xi)
& := & 
\left(\frac{R}{\sigma}\right)^{2}\Omega_{2}\left(\xi-
\frac{\phi(\xi)}{\bar{\Phi}(\xi)}\right)-\left[2 -
\gamma(\Omega_{1}-\Omega_{2}-\gamma)-\frac{\Omega_{2}R}{\sigma}\right]. \\
\end{array} \right.				
\end{align}
Then, it follows that  
$(\widehat{\theta}_{\mbox{\tiny z,MLE}},
\widehat{\sigma}_{\mbox{\tiny z,MLE}}) 
\sim 
\mathcal{AN}
\left((\theta,\sigma),
\frac{1}{n}\bm{S}_{\mbox{\tiny {z,MLE}}}
\right),
$
where 
\begin{equation}
\label{eqn:MLEZDeltaMatrix2}
\bm{S}_{\mbox{\tiny{z,MLE}}}
=
\frac{\Lambda^{-1}}
{\bar{\Phi}(\gamma)\left(\psi_{1}\psi_{3}-\psi_{2}^{2}\right)}
\mathbf{D}
\begin{bmatrix}
-\psi_{3} & \sigma \psi_{2} \\[10pt]
\sigma \psi_{2} & -\sigma^{2}\psi_{1} 
\end{bmatrix}
\mathbf{D}',
\quad
\mathbf{D}
\ \mbox{is given by (\ref{eqn:MLEYDeltaMatrix2})}.
\end{equation}

\subsection{Composite Model} 
\label{sec:CompositeModel}

In this section we briefly revisit the composite 
lognormal-Pareto model where a single threshold 
value assumed to be applied uniformly to the whole 
data set as developed by \cite{MR2188658} and 
\cite{MR2347211} but for payments $Y$ and $Z$ 
insurance data structures.
More general version of composite lognormal-Pareto model
where the threshold can vary among observations has 
also been investigated by \cite{MR2842558}.
These models can guard against
\cite{MR3474025} applied the composite models in the well-known Norwegian fire data and considered the data truncation (policy deductible) in parameter estimation and risk management. Here, we review the structure of \cite{MR2347211} models and implement them with both policy deductible and policy limit for payment $Y$ and payment $Z$ scenarios.

Let $f_{1}$ and $f_{2}$ be lognormal and single parameter 
Pareto density function, respectively 
\begin{align*}
f_{1}(x)
& = 
\frac{1}{\sqrt{2\pi}\sigma x}
e^{-\frac{1}{2}\left(\frac{\log (x-w_{0}) 
-\theta}{\sigma}\right)^2}, \quad 0 \le w_{0} < x
\quad 
\mbox{and}
\quad 
f_{2}(x)
= 
\frac{\alpha x_{0}^{\alpha}}{x^{\alpha+1}},
\quad x> x_{0}. 
\end{align*}
Assume the ground-up claim severity random 
variable $X$ has the composite pdf
\begin{align}
\label{eqn:mPDF}
f(x) 
& = 
\begin{cases}
w \dfrac{f_{1}(x)}{F_{1}(x_{0})}; 
& \mbox{if} \quad w_{0} < x \le x_{0}, \\
(1-w) f_{2}(x); 
& \mbox{if} \quad x > x_{0}.
\end{cases}
\end{align}
where $0 \leq w \leq 1$ is a mixing weight, 
$f_{1}$ and $F_{1}$ are the pdf and cdf representing
the "moderate" claims, and $f_{2}$ is the pdf describing "large" claims.
The corresponding cdf of $X$ is given as
\begin{align}
\label{eqn:mCDF}
F(x) 
& = 
\begin{cases}
w \dfrac{F_{1}(x)}{F_{1}(x_{0})}; 
& \mbox{if} \quad w_{0} < x \le x_{0}, \\
w+(1-w) F_{2}(x); 
& \mbox{if} \quad x > x_{0}.
\end{cases}
\end{align}
Here, $-\infty <\theta<\infty, \sigma>0, \alpha>0$ and $x_{0}>0$ are unknown parameters.
The continuity and differentiability conditions at $x_{0}$ are 
\[ 
f(x_{0}-)
=
f(x_{0}+)
\quad 
\mbox{and}
\quad 
f'(x_{0}-)=f'(x_{0}+),
\]
which leads to the constraints for $x_{0}$ and $w$, and reduces the number of estimated parameters
\[
\theta
= 
\ln (x_{0} - w_{0})
-  
\alpha \sigma^{2}
\quad 
\mbox{and} 
\quad 
w
=
\dfrac{\sqrt{2\pi}\alpha \sigma 
\Phi(\alpha \sigma) 
e^{\alpha \sigma^2/2}}{\sqrt{2\pi}
\alpha \sigma \Phi(\alpha \sigma)
e^{\alpha \sigma^2/2}+1}.
\]

Consider policy deductible, limit, and coinsurance
factor to be $d$, $u$, and $c$, respectively,
known in advance such that $w_{0} <d<x_{0}<u<\infty$.
With this assumption, if a sample of size $n$ is known
either for payment $Y$ or for payment $Z$, then the 
corresponding truncated and/or censored sample
can be recovered via linear transformation as
$x = y/c+d$ or $x = z/c+d$. 
Thus, without loss of generality
we may consider $x_{1},\hdots,x_{n}$ be an $i.i.d.$ 
sample either from payment $Y$ or from payment $Z$ composite model. 
The parameters estimation of $x_{0}, \sigma,$ and $\alpha$ 
are then derived by the maximum likelihood criterion.  
By using (\ref{eqn:mPDF}) and (\ref{eqn:mCDF}), we have
\begin{itemize}
\item[(a)] 
For payment $Y$, the log-likelihood function is
\begin{align}
l_{y}
& =
\sum_{i=1}^{n}
\log 
\left( 
f_{1}(x_{i})
\right) 
\ID(x_{i} \leq x_{0})
+ 
\log 
\left( 
\frac{w}{F_{1}(x_{0})}
\right) 
\sum_{i=1}^{n}
\ID \{ x_{i} \leq x_{0} \} 
\nonumber \\
& \quad 
+ 
\sum_{i=1}^{n}
\log 
\left( 
f_{2}(x_{i})
\right) 
\ID \{ x_{0} <x_{i} < u \} 
+ 
\log [1-F_{2}(u)]
\sum_{i=1}^{n}
\ID \{ x_{i} = u \} 
\nonumber \\
& \quad 
+ 
\log (1-w)
\sum_{i=1}^{n}
\ID \{ x_{i} > x_{0} \}
- 
n 
\log 
\left( 
1-w \dfrac{F_{1}(d)}{F_{1}(x_{0})}
\right).
\end{align}

\item[(b)] 
For payment $Z$, the log-likelihood function becomes
\begin{align}
l_{z}
& = 
\log \left( F_{1}(d) \right)
\sum_{i=1}^{n}
\ID \{ x_{i} = d \}
+
\sum_{i=1}^{n}
\log \left( f_{1}(x_{i}) \right) 
\ID \{ d< x_{i} \leq x_{0} \}
\nonumber \\ 
& \quad 
+ 
\log 
\left( 
\frac{w}{F_{1}(x_{0})}
\right) 
\sum_{i=1}^{n}
\ID \{ x_{i} \leq x_{0} \}
+ 
\sum_{i=1}^{n}
\log \left( f_{2}(x_{i}) \right) 
\ID \{ x_{0} <x_{i} < u \} 
\nonumber \\ 
& \quad 
+ 
\log [1-F_{2}(u)]
\sum_{i=1}^{n}
\ID \{ x_{i}=u \}
+ 
\log (1-w) 
\sum_{i=1}^{n}
\ID \{ x_{i} > x_{0} \}.
\end{align}
\end{itemize}

For notational convenient, the composite lognormal-Pareto 
with $f_{1}$ lognormal and $f_{2}$ single parameter Pareto 
will be denoted by \verb|LNPaI|.
An alternative composite lognormal-Pareto model with $f_{1}$ 
lognormal and $f_{2}$ a generalized Pareto pdf given by 
\[
f_{2}(x)
=
\frac{\alpha (\lambda + x_{0})^{\alpha}}
{(\lambda+x)^{\alpha+1}}, \quad x>x_{0}, 
\]
will be denoted by \verb|LNGPD|.

\section{MWM}
\label{sec:MWM}

MWM estimators are derived by following the standard method-of-moments 
approach, but instead of standard moments we match sample and population 
{\em winsorized\/} moments (or their variants).
Similar to MTM estimator, the population winsorized moments also always exist. 
The following definition lists the formulas of sample and population winsorized 
moments for the payment-per-payment and payment-per-loss data scenarios.

\begin{defn}
Let us denote the sample and population winsorized 
moments as 
$\widehat{W}_{j}$ and $W_j(\boldsymbol{\theta})$, respectively.
If 
$w_{1:n} \leq \cdots \leq w_{n:n}$
is an ordered realization of variables 
\eqref{p1Ndata} or \eqref{p2NZdata} 
with qf denoted by $F^{-1}_V(v \, | \, \boldsymbol{\theta})$
with 
$
V 
\in 
\{Y,Z\} 
$, 
then the sample and population winsorized moments,
with the winsorizing proportions $a$ (lower),
$b$ (upper) and $\bar{b} = 1-b$, have 
the following expressions:
\begin{eqnarray}
\widehat{W}_{j} 
& = & \frac{1}{n} 
\left[ 
m_n \big[ h(w_{m_n+1:n}) \big]^j +
\sum_{i=m_n+1}^{n-m_n^*} \big[ h(w_{i:n}) \big]^j +
m_n^* \big[ h(w_{n - m_n^*:n}) \big]^j 
\right], 
\label{Ws}
\\[1ex]
W_j(\boldsymbol{\theta}) 
& = & 
a \big[ h(F_V^{-1}(a \, | \, \boldsymbol{\theta})) \big]^j +
\int_{a}^{\bar{b}} \big[ h(F_V^{-1}(v \, | \, \boldsymbol{\theta})) \big]^j \, dv 
+
b \big[ h(F_V^{-1}(\bar{b} \, | \, \boldsymbol{\theta})) \big]^j,
\label{Wp}
\end{eqnarray}
where $j = 1, \ldots, k$, the winsorizing proportions $a$, $b$ and 
function $h$ are chosen by the researcher.
Also, integers $m_n$ and 
$m_{n}^* ~ (0 \le m_n < n - m_n^* \le n)$ are 
such that $m_n/n \rightarrow a$ 
and $m_n^*/n \rightarrow b$ when $n \rightarrow \infty$.
In finite samples, the integers $m_n$ and $m_{n}^*$
are computed as $m_n = [n a]$ and 
$m_{n}^* = [n b]$, where $[\cdot]$ 
denotes the greatest integer part. 

Winsorized-estimators are found by matching sample 
winsorized-moments (\ref{Ws}) with 
population winsorized moments (\ref{Wp}) for 
$j = 1, \ldots, k$, and then solving 
the system of equations with respect
to $\theta_1, \ldots, \theta_k$.
The obtained solutions, which we denote by 
$
\widehat{\theta}_j 
=
g_j(\widehat{W}_{1}, \ldots, \widehat{W}_{k})$,
$1 \leq j \leq k$, are, by definition, the 
MWM-estimators of 
$\theta_1, \ldots, \theta_k$. 
Note that the functions $g_j$ are such that
$
\theta_j 
=
g_j(W_1(\boldsymbol{\theta}),
\ldots, W_k(\boldsymbol{\theta})).
$
\end{defn}

The asymptotic theory of MWM estimators as a general
class of $L$-statistics can be found in \cite{MR0203874}
and a more computationally efficient expressions for 
completely observed data scenarios have
been established by \cite{MR3758788} and is given
by Theorem \ref{thm:PPY_MWM_AS1}.

\bigskip

\begin{thm}
\label{thm:PPY_MWM_AS1}
Suppose an i.i.d. realization of variables 
\eqref{p1Ndata} or \eqref{p2NZdata}
has been generated by cdf 
$F_V(v \, | \, \boldsymbol{\theta})$ 
which 
depending upon the data scenario equals to cdf 
given by $F_{V}, V \in \{Y,Z\}$, respectively. 
Let 
\[
\widehat{\boldsymbol{\theta}}_{\mbox{\tiny W}} 
= 
\left(
\widehat{\theta}_1, \ldots, \widehat{\theta}_k 
\right)
= 
\left( 
g_1
\left(
\widehat{W}_{1}, \ldots, \widehat{W}_{k}
\right), \ldots,
g_k
\left(
\widehat{W}_{1}, \ldots, \widehat{W}_{k}
\right) 
\right)
\]
denote 
a winsorized-estimator of $\boldsymbol{\theta}$. 
Then
\begin{align}
\widehat{\boldsymbol{\theta}}_{\mbox{\tiny W}} 
\label{eqn:MWM_Asy_Dist1}
= 
\left( \widehat{\theta}_1, \ldots, \widehat{\theta}_k \right) 
~~is~~ 
{\cal{AN}}
\left( 
\big( \theta_1, \ldots, \theta_k \big), \, 
\frac{1}{n} \, 
\mathbf{D} \boldsymbol{\Sigma} \mathbf{D}'
\right),
\end{align}
where $\mathbf{D} := \big[ d_{ij} \big]_{i,j=1}^{k}$ 
is the Jacobian of the transformations 
$g_1, \ldots, g_k$ evaluated at 
$\big( W_1(\boldsymbol{\theta}), 
\ldots, 
W_k(\boldsymbol{\theta}) \big)$
and 
$\mathbf{\Sigma} := \big[ \sigma^2_{ij} \big]_{i,j=1}^{k}$
is the variance-covariance  matrix with the entries 
\begin{align}
\label{eq:mwmVB_var_cov}
\sigma^2_{ij} 
& =
\widehat{A}_{i,j}^{(1)} + \widehat{A}_{i,j}^{(2)} 
+ 
\widehat{A}_{i,j}^{(3)} + \widehat{A}_{i,j}^{(4)},
\end{align}
where the terms $\widehat{A}_{i,j}^{(m)}, \; m = 1, \ldots, 4$, are 
specified in \cite{MR3758788}, Lemma A.1.
\end{thm} 

The asymptotic performance of the newly designed 
estimators will be measured via asymptotic 
relative efficiency (ARE) with respect 
to MLE and for two parameter case it is defined as 
\citep[see, e.g.,][]{MR595165,MR1652247}:
\begin{equation} 
\label{eq:infinite_relative_efficiency_benchmark_MLE}
ARE(\mathcal{C}, MLE) 
=
\left( 
\dfrac{\mbox{det}
\left(\bm{\Sigma}_{\mbox{\tiny MLE}}\right)}
{\mbox{det}
\left(\bm{\Sigma}_{\mbox{\tiny $\mathcal{C}$}}\right)}
\right)^{1/2},
\end{equation}
where 
$
\bm{\Sigma}_{\mbox{\tiny MLE}}
$
and 
$
\bm{\Sigma}_{\mbox{\tiny $\mathcal{C}$}}
$
are the asymptotic variance-covariance matrices of the MLE and 
$\mathcal{C}$ estimators, respectively, and `det' stands for 
the determinant of a square matrix. 
The main reason why MLE 
should be used as a benchmark procedure is its optimal 
performance in terms of asymptotic variance (of course, 
with the usual caveat of ``under certain regularity conditions''), 
for more details we refer to \cite{MR595165} \S4.1.

\subsection{Payments {\em Y}}
\label{sec:PPY_MWM}

Due to the piece-wise nature of the
qf $F_{Y}^{-1}$ with the transition 
point 
$
s^{*} 
=
\frac{F_{X}(T)-F_{X}(t)}{1-F_{X}(t)},
$
there are three possible arrangements among
$s^{*}, \ a$, and $b$ and they are:

\medskip

Case I: 
$0 \leq a < s^{*} \leq 1-b \leq 1$ 
~(estimation based on observed and censored data).

Case II: 
$0 \leq a < 1-b \leq s^{*} \leq 1$ 
~(estimation based on observed data only).

Case III: 
$0 < s^{*} \leq a < 1-b \leq 1$ 
~(estimation based on censored data only).

\medskip 

\noindent 
The empirical estimate $\hat{s}_{\mbox{\tiny E}}^{*}$ 
of $s^{*}$ is given by:
\begin{align}
\label{eqn:defnSStar1}
\hat{s}_{\mbox{\tiny E}}^{*}
& :=
\frac{F_{n}(T)-F_{n}(t)}{1-F_{n}(t)}
=
n^{-1} \sum_{i=1}^n \ID \{ 0 < y_i < cR \},
\quad \mbox{where $F_{n}$ is the empirical cdf}.
\end{align}
Case II simply implies the estimation based on 
observed data only and this is the most reasonable 
case to be considered as mentioned by \cite{MR4263275}.
Thus in this paper we will proceed with Case II.

Next, let $y_{1},\ldots,y_{n}$ be an {\em i.i.d.} sample 
of normal payment-per-payment data defined by (\ref{p1Ndata})
with qf $F_{Y}^{-1}$. Then for $k = 1,2$, we have
\begin{align}
\widehat{W}_{y,k}
& = 
\displaystyle
\frac{1}{n}
\left[ 
m_{n} \,
\left[ 
h_{Y}{\left(y_{m_{n}+1:n}\right)}
\right]^{k}
+ \sum_{i=m_{n}+1}^{n-m_{n}^{*}}
\left[ 
h_{Y}(y_{i:n})
\right]^{k}
+ m_{n}^{*} \,
\left[ 
h_{Y}{\left(y_{n-m_{n}^{*}:n}\right)} 
\right]^{k}
\right]  
\nonumber \\[20pt]
& = 
\displaystyle
\frac{1}{n}
\left[ 
m_{n} \, 
\left( 
\dfrac{y_{m_{n}+1:n}}{c}+t
\right)^{k}
+ \sum_{i=m_{n}+1}^{n-m_{n}^{*}}
\left(\dfrac{y_{i:n}}{c}+t\right)^{k}
+ m_{n}^{*} \, 
\left( 
\dfrac{y_{n-m_{n}^{*}:n}}{c}+t
\right)^{k}
\right],
\label{eqn:PPYMWMHat2}
\end{align}
with $m_{n}/n \rightarrow a$ and $m_{n}^{*}/n \rightarrow b$. 
With Case II, choose
$m_{n}^{*} \geq \sum_{i=1}^{n}\ID\{y_{i} = cR\}$.
The corresponding population winsorized moments  
\eqref{Wp} with the qf defined by $F_{Y}^{-1}$ are given by:
\begin{eqnarray}
W_{y,k}(\bm{\theta}) 
& = &
a 
\left[ 
h_Y 
\left( 
F_{Y}^{-1}(a \, | \, \bm{\theta}) 
\right) 
\right]^{k}
+ \int_a^{\bar{b}} 
\left[ 
h_Y 
\left(
F_Y^{-1} (s \, | \, \bm{\theta}) 
\right)
\right]^{k} \, ds
+ b 
\left[
h_Y 
\left(
F_{Y}^{-1}(\bar{b} \, | \, \bm{\theta}) 
\right) 
\right]^{k} 
\nonumber \\ 
& = &
a
\left[ 
\theta 
+ \Delta_{a}
\right]^{k}
+ 
\int_{a}^{\bar{b}}
\Delta_{s}^{k} \, ds 
+ b
\left[ 
\theta 
+ \Delta_{\bar{b}}
\right]^{k} 
\nonumber \\
& = & 
\begin{cases}
\theta + \sigma c_{y,1}, 
& \mbox{for} \ k = 1; \\
\theta^{2} + 2 \theta \sigma c_{y,1} + \sigma^{2} c_{y,2}, 
& \mbox{for} \ k = 2,
\end{cases}
\end{eqnarray}
where for $k=1,2$;
\begin{eqnarray}
\label{eq:PPMWMNormalCkdefine}
c_{y,k} 
& \equiv &
c_{y, k}(\Phi,a,b,\gamma) 
=
a
\Delta_{a}^{k}
+ \int_{a}^{\bar{b}}
\Delta_{s}^{k} \, ds 
+ b
\Delta_{\bar{b}}^{k}.
\end{eqnarray}

It is important to mention here that
$c_{y, k}$ depends on the unknown parameters
but does not depend on the parameters to be estimated for 
completely observed sample data
\citep[see, e.g.,][]{MR3758788}.
Equating 
$
W_{y,k} 
= 
\widehat{W}_{y,k},
$
for $k = 1, 2$
yield the implicit system of equations to be solved for 
$\theta$ and $\sigma$:
\begin{align}
\label{eqn:Normal_MTMPPY_Eqns}
\left\{
\begin{array}{lll}
\theta 
& = &
\widehat{W}_{y,1} - c_{y, 1}\sigma 
=: 
g_{1}
\left(
\widehat{W}_{y,1},\widehat{W}_{y,2}
\right), \\[10pt]
{\sigma} 
& = &
\sqrt{
\left(
\widehat{W}_{y,2}
-\widehat{W}_{y,1}^{2}
\right) \big/
\left(
c_{y, 2}-c_{y, 1}^{2}
\right)} 
=: 
g_{2}
\left(
\widehat{W}_{y,1},\widehat{W}_{y,2}
\right).
\end{array}
\right.
\end{align}
The system of equations (\ref{eqn:Normal_MTMPPY_Eqns}) can 
be solved for $\widehat{\theta}_{\mbox{\tiny y,MWM}}$ and 
$\widehat{\sigma}_{\mbox{\tiny y,MWM}}$ by using an iterative 
numerical method with the initializing values 
\begin{align}
\label{eqn:Normal_MTMPPY_EqnsStart}
\sigma_{\mbox{\tiny start}} 
& =
\sqrt{\widehat{W}_{y,2} - \widehat{W}_{y,1}^{2}} 
\quad \text{and} \quad 
\theta_{\mbox{\tiny start}} 
= 
\widehat{W}_{y,1}.
\end{align}

From Theorem \ref{thm:PPY_MWM_AS1},
the entries of the variance-covariance
matrix $\bm{\Sigma}_{y}$ 
calculated using \eqref{eq:mwmVB_var_cov} are
\begin{align*}
\sigma_{11}^{2} 
& = 
\sigma^2
c_{y,1}^{*}, 
\quad 
\sigma_{12}^{2} 
= 
\sigma_{21}^{2} 
=
2\theta \sigma^2 c_{y,1}^{*}
+ 
2 \sigma^3c_{y,2}^{*}, 
\quad \mbox{and} \quad 
\sigma_{22}^{2} 
= 
4\theta^2 \sigma^2c_{y,1}^{*}
+ 8 \theta \sigma^3c_{y,2}^{*}
+ 4 \sigma^4c_{y,3}^{*},
\end{align*}
where the expressions for 
$c_{y,k}^{*}, \ k = 1,2,3$
are listed in Appendix \ref{sec:AppendixA}.
For $k = 1,2$; it follows that
\begin{align}
\label{eqn:PPDerCks1}
\left\{
\begin{array}{lll}
\dfrac{\partial c_{y,k}}{\partial \theta} 
& = &
-\dfrac{k\phi(\gamma)}{\sigma}
\left\{
\dfrac{a\bar{a}
\Delta_{a}^{k-1}}
{\phi
\left(\Delta_{a}\right)}
+\displaystyle\int_{a}^{\bar{b}}\dfrac{\bar{s}
\Delta_{s}^{k-1}}
{\phi\left(\Delta_{s}\right)}\,ds
+\dfrac{b^2\Delta_{\bar{b}}^{k-1}}
{\phi\left(\Delta_{\bar{b}}\right)} 
\right\}, \\[15pt]
\dfrac{\partial c_{y,k}}{\partial \sigma} 
&=&  
-\dfrac{k(t-\theta)\phi(\gamma)}{\sigma^2}
\left\{
\dfrac{a\bar{a}
\Delta_{a}^{k-1}}
{\phi\left(\Delta_{a}\right)}
+\displaystyle\int_{a}^{\bar{b}}\dfrac{\bar{s}
\Delta_{s}^{k-1}}
{\phi\left(\Delta_{s}\right)}\,ds
+\dfrac{b^2\Delta_{\bar{b}}^{k-1}}
{\phi\left(\Delta_{\bar{b}}\right)} 
\right\}.
\end{array}
\right.
\end{align}
For $k=1,2$, let us denote
\begin{equation*}
\theta_{W_{y,k}} 
:=
\left. \frac{\partial g_{1}}{\partial \widehat{W}_{y,k}}\right|_{(W_{y,1};W_{y,2})}
\quad \text{and} \quad
\sigma_{W_{y,k}} 
:= 
\left. \frac{\partial g_{2}}{\partial \widehat{W}_{y,k}}\right|_{(W_{y,1};W_{y,2})}.
\end{equation*}
Consider the following additional notations
used by \cite{MR4263275} but for different 
$c_{y,k}$ functions.
\begin{align}
\label{eqn:DefOfFs1}
\left\{
\begin{array}{lllllll}
f_{11}(\theta, \sigma) 
& := &
1+\sigma \frac{\partial c_{y,1}}{\partial \theta}, & & 
f_{12}(\theta, \sigma) 
& := & 
c_{y,1}+\sigma \frac{\partial c_{y,1}}{\partial \sigma}, \\[5pt]
f_{12}(\theta, \sigma) 
& := & 
c_{y,1}+\sigma \frac{\partial c_{y,1}}{\partial \sigma}, & &
f_{22}(\theta, \sigma) 
& := &
\frac{\partial c_{y,2}}
{\partial \sigma}-2c_{y,1}
\frac{\partial c_{y,1}}{\partial \sigma}.
\end{array}
\right.
\end{align}
By multivariate chain rule and for 
$j,k \ge 1$, we have
\begin{align}
\label{eqn:MCR1}
\dfrac{\partial c_{y,k}}{\partial W_{y,j}}
& = 
\dfrac{\partial c_{y,k}}{\partial \theta} 
\dfrac{\partial \theta}{\partial W_{y,j}}
+
\dfrac{\partial c_{y,k}}{\partial \sigma} 
\dfrac{\partial \sigma}{\partial W_{y,j}} 
= 
\dfrac{\partial c_{y,k}}{\partial \theta} 
\theta_{W_{y,j}}
+
\dfrac{\partial c_{y,k}}{\partial \sigma} 
\sigma_{W_{y,j}}.
\end{align}

\noindent
Finally, the entries of the Jacobian matrix $\bm{D}_{y}$, 
given by Theorem \ref{thm:PPY_MWM_AS1},
are found by implicitly differentiating the functions $g_{j}$ 
from equations (\ref{eqn:Normal_MTMPPY_Eqns}) with the help 
of equation (\ref{eqn:PPDerCks1}) and using equations 
\eqref{eqn:DefOfFs1} and \eqref{eqn:MCR1}, we have
\begin{align*}
d_{11} 
& =
\theta_{W_{y,1}} 
= 
\frac{1-f_{12}\sigma_{W_{y,1}}}{f_{11}} 
= 
\frac{1-f_{12}d_{21}}{f_{11}}, \\[10pt]
d_{12} 
& = 
\theta_{W_{y,2}}
=
-\frac{f_{12}\sigma_{W_{y,2}}}{f_{11}} 
=
-\frac{f_{12}d_{22}}{f_{11}}, \\[10pt]
d_{21} 
& =
\sigma_{W_{y,1}} 
=
-\frac{K
\left[
2f_{11}W_{y,1}(c_{y,2}-c_{y,1}^{2}) 
+ f_{21}(W_{y,2}-W_{y,1}^{2})
\right]}
{f_{11}(c_{y,2}
-c_{y,1}^{2})^{2}
+K(W_{y,2}-W_{y,1}^{2})(f_{11}f_{22}
-f_{12}f_{21})}, \\[10pt]
d_{22} 
& =
\sigma_{W_{y,2}} 
=
\frac{Kf_{11}(c_{y,2}-c_{y,1}^{2})}{f_{11}(c_{y,2}
-c_{y,1}^{2})^{2}+K(W_{y,2}
-W_{y,1}^{2})(f_{11}f_{22}-f_{12}f_{21})}, 
\end{align*}
where 
$
K 
:=
\frac{1}{2}
\sqrt{\frac{c_{y,2}
-c_{y,1}^{2}}{W_{y,2}
-W_{y,1}^{2}}}.
$
Consequently,
\begin{gather}
\begin{aligned}
\bm{S}_{\mbox{\tiny y,MWM}}
:= 
\bm{D}_{y} \bm{\Sigma}_{y} \bm{D}_{y}'
& =
\left [ \begin{matrix}
d_{11} & d_{12}\\[12 pt]
d_{21} & d_{22}
\end{matrix} \right]
\left[ \begin{matrix}
\sigma_{11}^{2} & \sigma_{12}^{2}\\[12 pt]
\sigma_{21}^{2} & \sigma_{22}^{2}
\end{matrix} \right]
\left[ \begin{matrix}
d_{11} & d_{21}\\[12 pt]
d_{12} & d_{22}
\end{matrix} 
\right].
\end{aligned}
\end{gather}
Hence the asymptotic result \eqref{eqn:MWM_Asy_Dist1}
becomes
\begin{equation}
\label{eqn:PP_MWMEstNormal}
(\widehat{\theta}_{\mbox{\tiny y,MWM}},
\widehat{\sigma}_{\mbox{\tiny y,MWM}})
\sim \mathcal{AN}
\left((\theta,\sigma),n^{-1}\bm{S}_{\mbox{\tiny y,MWM}}\right), \ 
\bm{S}_{\mbox{\tiny y,MWM}}
:= 
\bm{D}_{y}\bm{\Sigma}_{y} \bm{D}_{y}^{'}.
\end{equation}
From (\ref{eqn:MLEYDeltaMatrix2}) and 
(\ref{eqn:PP_MWMEstNormal}), 
it follows that
\begin{align}
\label{eqn:PPY_MTM_MLE_ARE}
\mbox{ARE}
\left( 
\left(\widehat{\theta}_{\mbox{\tiny y,MWM}},
\widehat{\sigma}_{\mbox{\tiny y,MWM}}\right),
\left(\widehat{\theta}_{\mbox{\tiny y,MLE}},
\widehat{\sigma}_{\mbox{\tiny y,MLE}}\right)
\right)
& =
\left(
{\mbox{det}\left(\bm{S}_{\mbox{\tiny y,MLE}}\right)}/
{\mbox{det}
\left(
\bm{S}_{\mbox{\tiny y,MWM}}
\right)}
\right)^{0.5}.
\end{align}
In Table \ref{table:PPY_MLE_TW_ARE}, we provide AREs 
of MTM and MWM estimators for selected trimming and 
winsorizing proportions $a$ and $b$.

\begin{table}[hbt!]
\centering
\caption{
$
\mbox{ARE}
\left( 
\left(
\widehat{\theta}_{\mbox{\tiny y,$\mathcal{C}$}},
\widehat{\sigma}_{\mbox{\tiny y,$\mathcal{C}$}}
\right),
\left(
\widehat{\theta}_{\mbox{\tiny y,MLE}},
\widehat{\sigma}_{\mbox{\tiny y,MLE}}
\right)
\right)
$
with 
$
\mathcal{C} 
\in 
\{ \mbox{MTM, MWM} \}
$ 
for fixed $d = 3$ and selected $a$ and $b$ and various choices of
right-censoring point $u$ from $\mbox{LN}(1,4,2)$.
}
\label{table:PPY_MLE_TW_ARE}
{\small 
\begin{tabular}{c|c|ccccc|cccc|ccc}
\cline{2-14}
{} & 
\multirow{2}{*}{$a$} &
\multicolumn{5}{c}{$b$ (when $u=5.96\times10^{3}$)} &
\multicolumn{4}{|c}{$b$ (when $u=1.54\times10^{3}$)} &
\multicolumn{3}{|c}{$b$ (when $u=7.52\times10^{2}$)} \\[-0.5ex]
\cline{3-14}
{} & 
{} & 
0.01 & 0.05 & 0.10 & 0.15 & 0.25 & 
0.05 & 0.10 & 0.15 & 0.25 & 0.10 & 0.15 & 0.25 \\
\hline
\hline
\multirow{5}{*}{\rotatebox{90}{MWM}} &
0 & 1.000 & 0.950 & 0.892 & 0.835 & 0.724 & 
1.000 & 0.938 & 0.878 & 0.762 & \fbox{0.999} & 0.936 & 0.811 \\
&0.05 & 0.995 & 0.945 & 0.886 & 0.829 & 0.718 & 
0.994 & 0.932 & 0.872 & 0.755 & 0.993 & 0.929 & 0.804 \\
&0.10 & 0.982 & 0.932 & 0.873 & 0.816 & 0.704 & 
0.981 & 0.919 & 0.858 & 0.741 & 0.978 & 0.914 & 0.789 \\
&0.15 & 0.963 & 0.913 & 0.853 & 0.796 & 0.684 & 
0.960 & 0.898 & 0.837 & 0.719 & 0.956 & 0.892 & 0.766 \\
&0.25 & 0.907 & 0.856 & 0.796 & 0.738 & 0.626& 
0.901 & 0.838 & 0.777 & 0.658 & 0.892 & 0.828& 0.701 \\
\hline
\multirow{5}{*}{\rotatebox{90}{MTM}} & 0 & 0.990 & 0.917 & 0.841 & 0.772 & 0.650 & 
0.964 & 0.884 & 0.813 & 0.684 & \fbox{0.942} & 0.866 & 0.728 \\
&0.05 & 0.983 & 0.913 & 0.839 & 0.772 & 0.652 & 
0.960 & 0.882 & 0.813 & 0.686 & 0.940 & 0.866 & 0.731 \\
&0.10 & 0.961 & 0.894 & 0.823 & 0.758 & 0.641 & 
0.941 & 0.865 & 0.797 & 0.674 & 0.922 & 0.850 & 0.718 \\
&0.15 & 0.930 & 0.866 & 0.797 & 0.734& 0.619 & 
0.911 & 0.838 & 0.772 & 0.652 & 0.893 & 0.823 & 0.694 \\
&0.25 & 0.854 & 0.795 & 0.730 & 0.670 & 0.560 & 
0.836 & 0.768 & 0.705& 0.589 & 0.818 & 0.751 & 0.628 \\
\hline
\end{tabular}
}
\end{table}

\noindent
Several important conclusions emerge from 
Table \ref{table:PPY_MLE_TW_ARE}.

\begin{itemize}
\item 
For lognormal model, the efficiency of MWM and MTM estimators
depends on the location of the right-censoring point. 
In particular, as the censoring point decreases the ARE values 
increase remarkably. 
For example, when $(a, b) = (0.05, 0.15)$,
$u=5.96\times10^{3}$  has
$\text{ARE}= 0.829; u=1.54\times10^{3}$ has $\text{ARE}= 0.872$; 
and $u=7.52\times10^{2}$
has $\text{ARE}= 0.929$. This can be explained by the fact that 
as the right censored point is closer to the policy limit, lower 
volume of data are affected by the censoring structure and larger 
efficiency are kept in values of ARE. 

\item
It is also of interest to compare the MWM approach with the MTM 
for various right-censoring points.
Table \ref{table:PPY_MLE_TW_ARE} contains ARE entries for 
the MTM estimators, which are taken from
\cite{MR2497558} and \cite{MR4263275}.
We clearly see that MWM uniformly dominates MTM in terms of ARE, 
while still offering identical breakdown points (degrees of 
resistance against lower and upper outliers) and computational 
efficiency. Furthermore, the benefit of MWM tends to be more 
significant when the policy limit becomes smaller. 

\item Besides the degree of efficiency, loss is different from 
MWM to MTM when the chosen proportions $a$ and $b$ vary. 
For example, with a fixed $b=0.10$, ARE of MTM approximately 
decreases by 2\% from $a=0.05$ to $a=0.10$ while ARE of MWM 
decrease by 1.4\% no matter what the value of $u$. This indicates 
the relative stability of MWM compared to MTM when the policy 
deductible and limit are considered in payment-per-payment cases.
\end{itemize}

\subsection{Payments {\em Z}}
\label{sec:PPZ_MWM}

From (\ref{p2NZdata}), it follows that payment $Z$ is a
left- and right-censored version of random variable $X$.
Thus, possible permutations between $a$, $b$, and their 
positioning with respect to $F(t)$ and $F(T)$ have to be 
taken into account since the expressions for $\sigma_{ij}^{2}$ 
given by \eqref{eq:mwmVB_var_cov} with qf $F_{Z}^{-1}$ depend
on the six possible permutations among $a$, $b$,
$F(t)$, and $F(T)$.
\begin{enumerate}
\item $0 \leq a < \bar{b} \leq F(t) < F(T) \leq 1$. \qquad \qquad 
4. $0 \leq F(t) < F(T) \leq a < \bar{b} \leq 1$.
\item $0 \leq a \leq F(t) < \bar{b} \leq F(T) \leq 1$. \qquad \qquad 
5. $0 \leq F(t) \leq a < F(T) \leq \bar{b} \leq 1$.
\item $0 \leq a \leq F(t) < F(T) \leq \bar{b} \leq 1$. \qquad \qquad 
6. $0 \leq F(t) \leq a < \bar{b} \leq F(T) \leq 1$.
\end{enumerate}
Among the six cases, two of those scenarios (estimation based 
on censored data only -- Cases 1 \& 4) have no parameters to be 
estimated in the formulas of population winsorized moments and three 
(estimation based on observed and censored data -- Cases 2, 3, \& 5)
are inferior to the estimation scenario based on fully observed data. 
Thus, from now on we will proceed only with Case 6 which makes
most sense and simplifies the estimation procedure significantly 
because it uses the available data in the most effective way. 
Moreover, the MWM-estimators based on Case 6 will be resistant 
to outliers, i.e., observations that are inconsistent with the 
assumed model and most likely appearing at the boundaries $t$ 
and $T$. Case 6 also eliminates heavier point masses given at 
the censored points $t$ and $T$. 

\bigskip

For practical data analysis purposes, standard empirical estimates 
of $F(t)$ and $F(T)$ provide guidance about the choice of $a$ and 
$\bar{b}$ and are chosen according to
\begin{equation}
\label{eqn:PPZ_Cond1}
F_{n}(t) 
\le a 
< \bar{b} 
\le F_{n}(T), 
\quad \mbox{where} \quad F_{n} \ 
\mbox{is the empirical cdf.}
\end{equation}
Further, define 
$h_Z(z) := z/c+t$. 
Now, consider an observed {\em i.i.d.\/} sample
$z_{1},\ldots,z_{n}$ defined by 
$F_{Z}^{-1}$.
Let $z_{1:n}, \ldots, z_{n:n}$ be the corresponding 
order statistics. 
Then the sample winsorized moments,
for $k = 1,2$, are given by 
(\ref{Ws})
\begin{align}
\widehat{W}_{z,k}
& = 
\displaystyle
\frac{1}{n}
\left[ 
m_{n} \,
\left[ 
h_{Z}{
\left(
z_{m_{n}+1:n}
\right)}
\right]^{k}
+ \sum_{i=m_{n}+1}^{n-m_{n}^{*}}
\left[ 
h_{Z}(z_{i:n})
\right]^{k}
+ m_{n}^{*} \,
\left[ 
h_{Z}
{
\left(
z_{n-m_{n}^{*}:n}
\right)
} 
\right]^{k}
\right]  
\nonumber \\[20pt]
& = 
\displaystyle
\frac{1}{n}
\left[ 
m_{n} \, 
\left( 
\dfrac{z_{m_{n}+1:n}}{c}
+ t
\right)^{k}
+ \sum_{i=m_{n}+1}^{n-m_{n}^{*}}
\left(\dfrac{z_{i:n}}{c}+t\right)^{k}
+ m_{n}^{*} \, 
\left( 
\dfrac{z_{n-m_{n}^{*}:n}}{c}+t
\right)^{k}
\right],
\label{eqn:PPZMWMHat2}
\end{align}
with $m_{n}/n \rightarrow a$ and $m_{n}^{*}/n \rightarrow b$. 
With Case 6, choose
$m_{n} \geq \sum_{i=1}^{n}\ID\{z_{i} = 0\}$ and 
$m_{n}^{*} \geq \sum_{i=1}^{n}\ID\{z_{i} = cR\}$.
By assuming the most general case that 
$0 \leq F(t) \leq a < \bar{b} \leq F(T) \leq 1$,
the corresponding population winsorized moments
\eqref{Wp}
with the qf $F_{Z}^{-1}$ are given by:
\begin{align}
W_{z,k}(\bm{\theta}) 
& = 
a 
\left[ 
h_Z 
\left( 
F_{Z}^{-1}(a \, | \, \bm{\theta}) 
\right) 
\right]^{k}
+ \int_a^{\bar{b}} 
\left[ 
h_Z 
\left(
F_Z^{-1} (s \, | \, \bm{\theta}) 
\right)
\right]^{k} \, ds
+ b 
\left[
h_Z 
\left(
F_{Z}^{-1}(\bar{b} \, | \, \bm{\theta}) 
\right) 
\right]^{k} 
\nonumber \\
& = 
a
\left[ 
F^{-1}(a \, | \, \bm{\theta}) 
\right]^{k}
+ \int_{a}^{\bar{b}}
\left[ 
{F^{-1}(s \, | \, \bm{\theta})}
\right]^{k} \, ds 
+ b
\left[ 
F^{-1}(\bar{b} \, | \, \bm{\theta})
\right]^{k} 
\nonumber \\
& =
\begin{cases}
\theta + \sigma c_{1}, 
& \mbox{for} \ k = 1; \\
\theta^{2} + 2 \theta \sigma c_{1} + \sigma^{2} c_{2}, 
& \mbox{for} \ k = 2,
\end{cases}
\label{eqn:Z_Pop_Mean}
\end{align}
where $c_{k} \equiv c_{y,k}, \ 1 \le k \le 4$,
given by (\ref{eq:PPMWMNormalCkdefine})
with $\gamma=-\infty$ 
and are listed in Appendix \ref{sec:AppendixA}.
Thus, with the assumption 
$0 \leq F(t) \leq a < \bar{b} \leq F(T) \leq 1$, 
this case translates to the complete data
case which is fully investigated by \cite{MR3758788}
and the MWM estimators of $\theta$ and $\sigma$ are
\begin{align}
\label{eqn:PPY_MWM_Estimators1}
\left\{
\begin{array}{rcl}
\widehat{\theta}_{\mbox{\tiny z,MWM}} 
& = & 
\widehat{W}_{z,1}
-c_{1} \, \widehat{\sigma}_{\mbox{\tiny z,MWM}} \\[5pt]
\widehat{\sigma}_{\mbox{\tiny z,MWM}}
& = & 
\sqrt{
	\left(\widehat{W}_{z,2} - \widehat{W}_{z,1}^{2}\right)/
	\left(c_{2}-c_{1}^{2}\right)}
\end{array}
\right.
\end{align}
The corresponding ARE is given by:
\begin{align}
\label{eqn:PPZ_MTM_MLE_ARE}
\mbox{ARE}
\left( 
\left(\widehat{\theta}_{\mbox{\tiny z,MWM}},
\widehat{\sigma}_{\mbox{\tiny z,MWM}}\right),
\left(\widehat{\theta}_{\mbox{\tiny z,MLE}},
\widehat{\sigma}_{\mbox{\tiny z,MLE}}\right)
\right)
& =
\left(
{\mbox{det}\left(\bm{S}_{\mbox{\tiny z,MLE}}\right)}/
{\mbox{det}
\left(
\bm{S}_{\mbox{\tiny z,MWM}}
\right)}
\right)^{0.5},
\end{align}
where 
\begin{align}
\label{eqn:SzMWM}
\bm{S}_{\mbox{\tiny z,MWM}}
& := 
\dfrac{\sigma^2}{\left(c_{2}-c_{1}^{2}\right)^{2}}
\begin{bmatrix}
c_{1}^{*}c_{2}^{2}-2c_{1}c_{2}c_{2}^{*}+c_{1}^{2}c_{3}^{*} & 
-c_{1}^{*}c_{1}c_{2}+c_{2}c_{2}^{*}+c_{1}^{2}c_{2}^{*}-c_{1}c_{3}^{*} \\[5pt]
-c_{1}^{*}c_{1}c_{2}+c_{2}c_{2}^{*}+c_{1}^{2}c_{2}^{*}-c_{1}c_{3}^{*} & 
c_{1}^{*}c_{1}^{2}-2c_{1}c_{2}^{*}+c_{3}^{*} 
\end{bmatrix},
\end{align}
where the expressions for 
$c_{k}^{*}, \ k = 1,2,3$, as functions of 
$a,b,c_{1},c_{2},c_{3}$, and $c_{4}$ 
are such that
$
c_{k}^{*}
\equiv 
c_{y,k}^{*}
$
with $\gamma = -\infty$
and are listed in Appendix \ref{sec:AppendixA}.

\begin{table}[tbh!]
\centering
\caption{$
\mbox{ARE}
\left( 
\left(
\widehat{\theta}_{\mbox{\tiny z,$\mathcal{C}$}},
\widehat{\sigma}_{\mbox{\tiny z,$\mathcal{C}$}}
\right),
\left(
\widehat{\theta}_{\mbox{\tiny y,MLE}},
\widehat{\sigma}_{\mbox{\tiny y,MLE}}
\right)
\right)
$
with 
$
\mathcal{C} 
\in 
\{ \mbox{MTM, MWM} \}
$ 
for fixed $d = 3$ and selected $a$ and $b$ and various choices of
right-censoring point $u$ from $\mbox{LN}(1,4,2)$.
}
\label{table:PPZ_MLE_TW_ARE}
{\small 
\begin{tabular}{c|c|ccccc|cccc|ccc}
\cline{2-14}
\multirow{2}{*}{} &
\multicolumn{1}{c}
{\multirow{2}{*}{$a$}}&
\multicolumn{5}{|c}{$b$ (when $u=5.96\times10^{3}$)} &
\multicolumn{4}{|c}{$b$ (when $u=1.54\times10^{3}$)} &
\multicolumn{3}{|c}{$b$ (when $u=7.52\times10^{2}$)} \\[-0.5ex]
\cline{3-14}
{} & 
{} & 0.01 & 0.05 & 0.10 & 0.15 & 0.25 & 
0.05 & 0.10 & 0.15 & 0.25 & 0.10 & 0.15 & 0.25 \\
\hline
\hline
\multirow{4}{*}{\rotatebox{90}{MWM}} &
0.10 & 0.954& 0.927 & 0.891 & 0.854 & 0.776 & 
0.961 & 0.923 & 0.884& 0.804& 0.965 & 0.924 & 0.840 \\
&0.15 & 0.896 & 0.867 & 0.830 & 0.791 & 0.711 & 
0.899 & 0.860 & 0.819 & 0.737 & 0.899 & 0.857& 0.770 \\
&0.25 & 0.795 & 0.765 & 0.726 & 0.686 & 0.603& 
0.793 & 0.752& 0.710& 0.625 & 0.786 & 0.743 & 0.653 \\
&0.49 & 0.570 & 0.538 & 0.495 & 0.452 & 0.363 & 
0.557 & 0.513 & 0.468 & 0.376 & 0.536 & 0.490& 0.393 \\
\hline
\multirow{4}{*}{\rotatebox{90}{MTM}} &
0.10 & 0.909 & 0.863 & 0.810 & 0.761 & 0.667 & 
0.894 & 0.839 & 0.788 & 0.690 & 0.878& 0.824 & 0.722 \\
&0.15 & 0.855 & 0.812 & 0.761 & 0.712& 0.621 & 
0.841 & 0.788 & 0.738 & 0.643 & 0.824& 0.771 & 0.672 \\
&0.25 & 0.754 & 0.714 & 0.667 & 0.621 & 0.534 & 
0.740 & 0.690 & 0.643 & 0.553 & 0.722 & 0.672 & 0.578 \\
&0.49 & 0.528 & 0.495 & 0.452 & 0.411& 0.329 & 
0.512 & 0.469 & 0.426 & 0.340 & 0.490 & 0.445 & 0.356 \\
\hline
\end{tabular}
}
\end{table}

\noindent
Similar to Table \ref{table:PPY_MLE_TW_ARE}, several important 
conclusions can be made based on Table \ref{table:PPZ_MLE_TW_ARE}.
\begin{itemize}
\item 
Similar to payments $Y$ (i.e., Table \ref{table:PPY_MLE_TW_ARE}),
MWM estimators consistently outperform MTM estimators in terms of ARE. 
The choice of right-censoring point (limit the proportion of complete 
data) significantly affects the level of model efficiency.
\item 
Unlike the payment $Y$ scheme, the degree of efficiency loss 
(with complete data) is almost identical from the MWM approach 
to the MTM method. This can be explained by the fact that payment 
$Z$ is an unconditional variable, and the density function used for 
both MTM and MWM procedures is not restricted by the conditional 
status of policy deductible and limit. Thus, on such a payment-per-loss 
base, the ARE results only depend on the fraction of $a$ and $b$ 
in this lognormal model.
\end{itemize}

\begin{figure}[hbt!]
\centering
\includegraphics[width=0.48\textwidth]{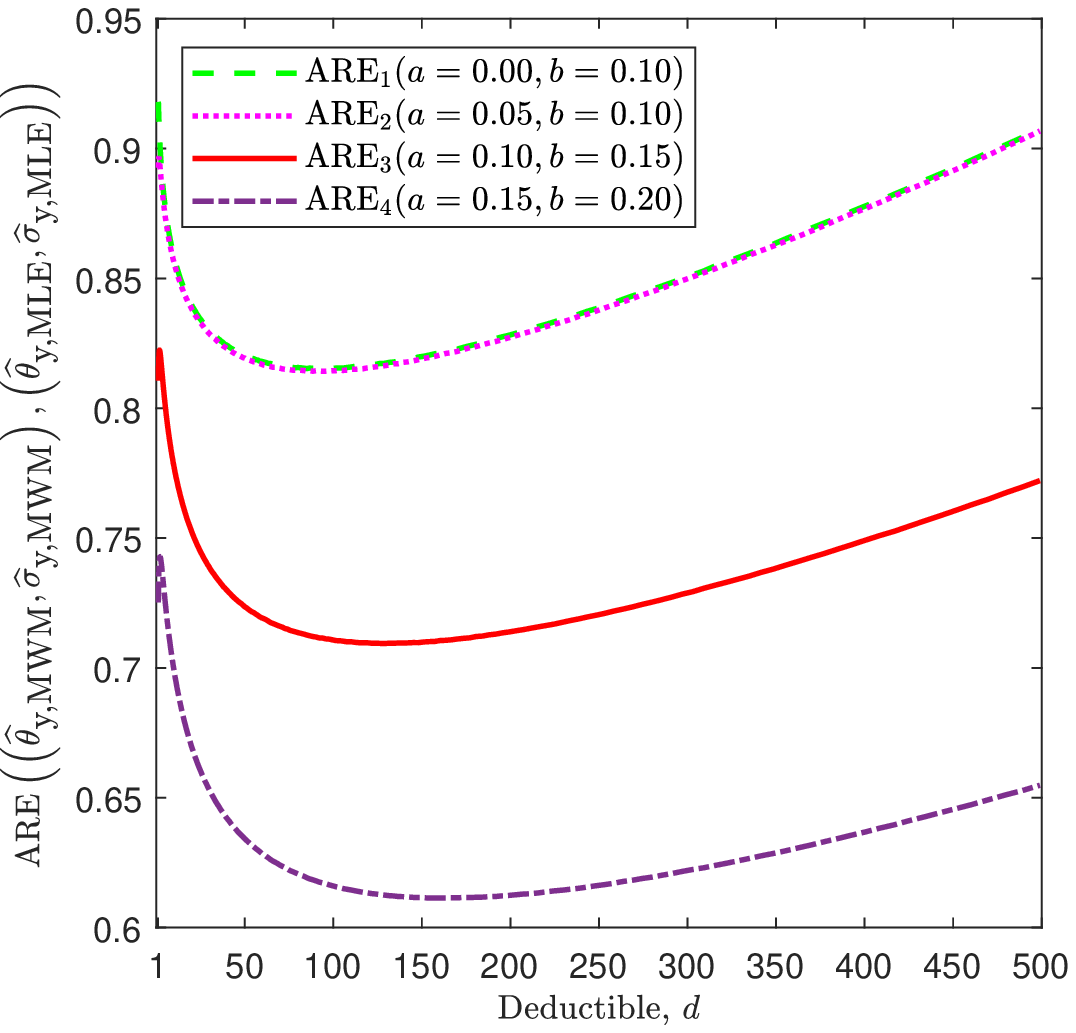}
\includegraphics[width=0.48\textwidth]{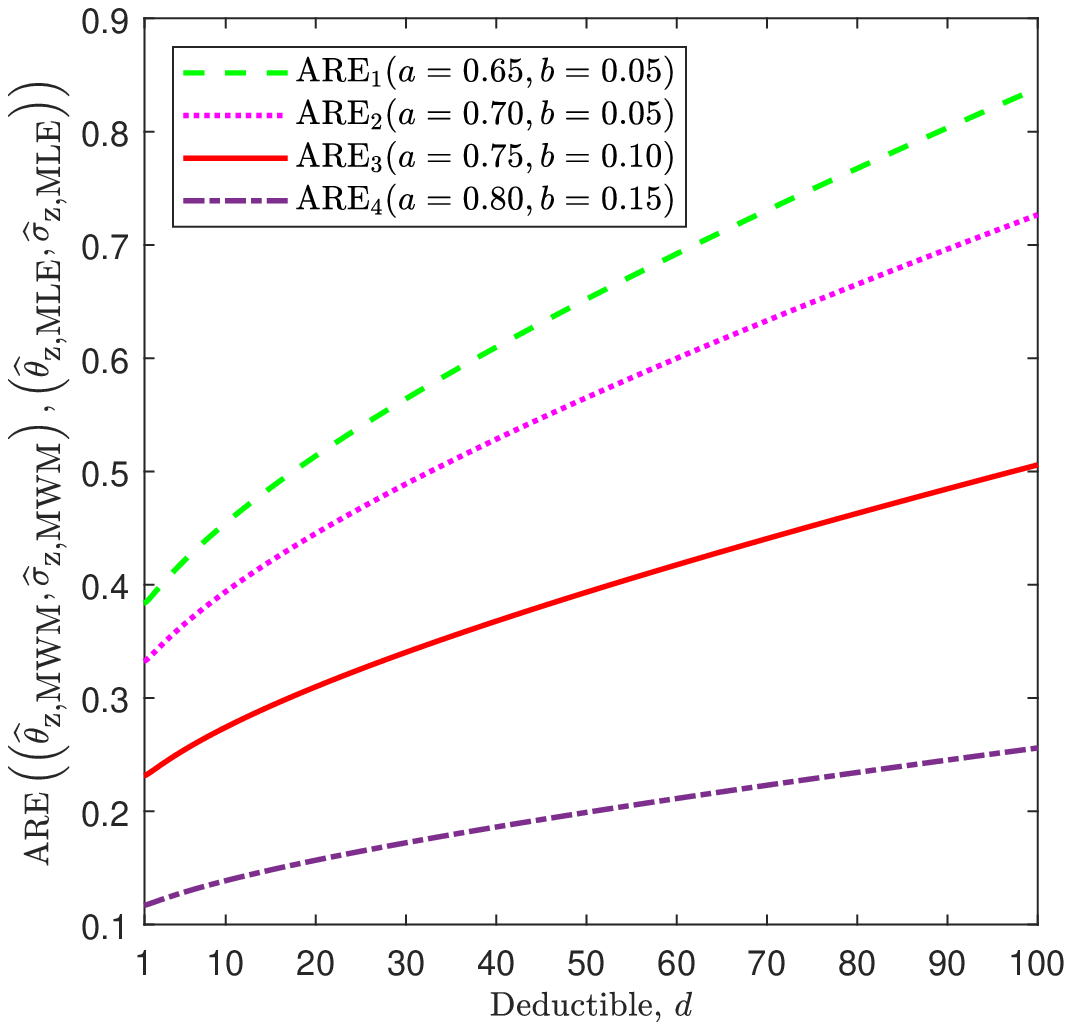}
\caption{
$
\mbox{ARE}
\left(
\left(
\widehat{\theta}_{\mbox{\tiny v,MWM}},\widehat{\sigma}_{\mbox{\tiny v,MWM}}
\right),
\left(
\widehat{\theta}_{\mbox{\tiny v,MLE}},\widehat{\sigma}_{\mbox{\tiny v,MLE}}
\right)
\right)$
curves as functions of policy deductible $d$ 
and with fixed policy limit $u = 5.96 \times 10^{3}$,
where $v = y$ (left panel) and $v = z$ (right panel),
respectively, represent the payments $Y$ and $Z$ 
data scenarios from 
$
\mbox{LN}
(1,4,2)
$
distribution.
\label{fig:ARE_MWM_Curves}
}
\end{figure}

Clearly the ARE expressions
given by Eqs. \eqref{eqn:PPY_MTM_MLE_ARE},
and \eqref{eqn:PPZ_MTM_MLE_ARE}
are functions of $d$, $u$, the parameters to be estimated, 
and the trimming and wisnorizing proportions. 
It is important to note here that those AREs 
for completely observed ground-up loss 
data set do not depend on the parameters 
to be estimated, see, e.g., \cite{MR2497558}
and \cite{MR3758788}. 
As functions of policy deductible, $d$, 
those AREs are visualized in Figure 
\ref{fig:ARE_MWM_Curves} for some selected
winsorizing proportions satisfying 
Case III as specified in Section \ref{sec:PPY_MWM}
for payment $Y$ and Case 6 as considered in 
Section \ref{sec:PPZ_MWM} for payment $Z$.

For payment $Y$ (Figure \ref{fig:ARE_MWM_Curves}, left panel),
the ARE curves for the left winsorizing proportions 
$a \in \{0.00, 0.05\}$
with fixed right winsorizing proportion $b = 0.10$  
do not have significant difference, but when
the left winsorizing proportion is increasing
then the ARE values are going down.
Since the ARE curves do not have consistent 
patter (either increasing or decreasing) 
as functions of $d$, this makes the practical
interpretation of those curves more difficult. 

On the other hand, for payment $Z$,
the ARE curves are consistently 
increasing functions of $d$ 
(Figure \ref{fig:ARE_MWM_Curves}, right panel).
This is because higher the value of $d$ implies
the narrower support for exact loss observations 
(which are bigger than $d$ and smaller than $u$)
which eventually yields the smaller variance of
MWM estimators.

\section{Simulation Study}
\label{sec:SimStudy}

This section supplements the theoretical results 
we developed in Section \ref{sec:MWM} via simulation. 
The main goal is to assess the size of the sample such
that the finite estimators are free from bias 
(given that the estimators are asymptotically unbiased),
justify the asymptotic normality, and
their finite sample relative efficiencies (REs)  
to reach the corresponding AREs. 
To compute RE of MWM estimators 
we use MLE as a benchmark. 
Thus, the definition of ARE given by equation 
(\ref{eq:infinite_relative_efficiency_benchmark_MLE})
for finite sample performance translates to:
\begin{equation} 
\label{eq:finite_relative_efficiency_benchmark_MLE}
RE(\mbox{MWM, MLE}) 
=
\frac{\text{asymptotic variance of MLE estimator}}
{\text{small-sample variance of a competing MWM estimator}},
\end{equation}
where the numerator is as defined in 
(\ref{eq:infinite_relative_efficiency_benchmark_MLE})
and the denominator is given by:
\[
\left(
\mbox{det}
\begin{bmatrix}
E 
\left[ 
\left(\widehat{\theta} - \theta \right)^{2}
\right] & 
E 
\left[ 
\left(\widehat{\theta} - \theta \right)
\left(\widehat{\sigma} - \sigma \right)
\right] \\[15pt]
E 
\left[ 
\left(\widehat{\theta} - \theta \right)
\left(\widehat{\sigma} - \sigma \right)
\right] & 
E 
\left[ 
\left(\widehat{\sigma} - \sigma \right)^{2}
\right] 
\end{bmatrix}
\right)^{1/2}.
\]

\subsection{Study Design}

From a ground-up lognormal distribution 
$F_{W}(w_{0}=1, \theta=4,\sigma=2)$, 
we generate 10,000 samples of a specified length $n$ using
Monte Carlo. 
For each sample we estimate the parameters of $F$ using various
MTM and MWM estimators and then compute the average mean and 
RE of those 10,000 estimates. 
The standardized ratio $\hat{\theta}/\theta$ that we report 
is defined as the average of 10,000 estimates divided by 
the true value of the parameter that we
are estimating. The standard error is standardized in a 
similar manner.

\begin{table}[hbt!]
\centering
\caption{Selection of trimming and winsorizing proportions.}
\label{table:MTM_MWM_Spe}
\begin{tabular}{c|c|c|c|c|c|c|c|c}
\hline 
\multicolumn{2}{c|}{Method} & 
\multicolumn{1}{|c|}{MLE} & 
\multicolumn{6}{|c}{MTM/MWM} \\
\hline 
\multirow{2}{*}{Payments $Y$} & 
$a$ & {--} & 
0.05 & 0.10 & 0.15 & 0.00 & 0.00 & 0.00 \\
\cline{2-9}
{} & $b$ & {--} & 
0.05 & 0.10 & 0.15 & 0.05 & 0.10 & 0.25 \\
\hline 
\multirow{2}{*}{Payments $Z$} & 
$a$ & {--} & 
0.10 & 0.15 & 0.25 & 0.10 & 0.10 & 0.10 \\
\cline{2-9}
{} & $b$ & {--} & 
0.10 & 0.15 & 0.25 & 0.05 & 0.15 & 0.25 \\
\hline 
\end{tabular}
\end{table}

We observe the performance of different methods of estimation 
for lognormal distribution. 
The simulation study is designed 
with following parameters:

\begin{enumerate}[label=(\roman*)]
\setlength{\itemsep}{-0.5em}
\item {\em Sample size\/}: 
$n = 100, \, 500, \, 1000, \, \infty$

\item
{\em Coinsurance rate\/}: 
$c = 1$.

\item 
{\em Truncation and censoring thresholds 
for both variables $Y$ and $Z$:}
\begin{itemize}
\item 
$d = 3$ 
(corresponding to about 5\% left-truncation); 
\item 
$u_{1} = 5.96 \times 10^3$
(corresponding to about 1\% right-censoring);
\item 
$u_{2} = 1.54 \times 10^3$
(corresponding to about 5\% right-censoring).
\end{itemize}

\item 
{\em Selection of trimming and winsorizing proportions:}
\begin{itemize}
\item 
As mentioned in Section \ref{sec:PPY_MWM},
for payments $Y$, the trimmed- and 
winsorized-estimators are derived under the condition 
that $0 \le a < 1-b \le s^{*}$
(i.e., Case II).
In simulations, however, the proportion $s^{*}$ is random, which
can easily result in violations of the specified condition. If 
$s^*$ is estimated empirically, i.e., by replacing $F$ with its 
empirical estimator $\widehat{F}_n$, then the variance of such 
estimator is equal to
$
\sigma_{\hat{s}^{*}}^{2}
=
\frac{{s}^{*}\left(1-{s}^{*}\right)}{n}
$.
Thus, to minimize the number of possible violations, the right 
trimming proportion $b$ is chosen to satisfy the following inequality:
\begin{align}
\label{eqn:CondsStarB}
1-b 
& \le 
s^{*} - 2\sigma_{\hat{s}^{*}},
\end{align}
and, to demonstrate the difference between sufficiently and 
insufficiently robust trimmed- and winsorized-estimators, some values 
of $b$ are chosen to violate the condition
$
1-b \le s^{*} - 2\sigma_{\hat{s}^{*}}.
$

\item 
Again, 
as mentioned in Section \ref{sec:PPZ_MWM}
and for payments $Z$, the equivalent condition is
$F(d) \le a < 1-b \le F(u)$. Similar arguments as those for 
payments $Y$ lead to
$\sigma_{\hat{F}_{n}(d)}^{2} = \frac{F(d)(1-F(d))}{n}$
and
$\sigma_{\hat{F}_{n}(u)}^{2} = \frac{F(u)(1-F(u))}{n}$.
In this case, 
the left and right trimming/winsorizing proportions $a$ and $b$, 
respectively, are chosen such that
\[
F(d) + 2 \sigma_{\hat{F}_{n}(d)} \le a
\quad 
\mbox{and}
\quad 
1-b \le F(u) - 2 \sigma_{\hat{F}_{n}(u)}.
\]
\end{itemize}

\item 
{\em Estimators of $\theta$ and $\sigma$\/}: 
MLE, MTM, and MWM estimators with the trimming 
and Winsorizing proportions as specified in 
Table \ref{table:MTM_MWM_Spe}.
\end{enumerate}

\begin{table}[hbt!]
\caption{
Lognormal payment-per-payment 
actuarial loss scenario,
$LN(w_{0}=1,\theta=4,\sigma=2)$ with $d = 3$
and two selected values of right-censoring point $u$.
The entries are mean values 
based on 10,000 samples.
}
\label{table:PPY_SimStudy_TW}
\centering
{\scriptsize 
\begin{tabular}{c|cc|cccc|cccc|cccc|cccc}
\cline{2-19}
{} & 
\multicolumn{2}{c|}{Proportion} & 
\multicolumn{4}{|c|}{$n=100$} & 
\multicolumn{4}{|c|}{$n=500$} &
\multicolumn{4}{|c|}{$n=1000$} &
\multicolumn{4}{|c}{$n \to \infty$} \\
\cline{2-19} 
& & & & & & & & & 
& & & & & & & & & \\[-2.50ex]
{} & 
\multirow{2}{*}{$a$} & 
\multirow{2}{*}{$b$} & 
\multicolumn{2}{|c}{\sc MWM} & 
\multicolumn{2}{c|}{\sc MTM} & 
\multicolumn{2}{c}{\sc MWM} & 
\multicolumn{2}{c|}{\sc MTM} &
\multicolumn{2}{c}{\sc MWM} & 
\multicolumn{2}{c|}{\sc MTM} &
\multicolumn{2}{c}{\sc MWM} & 
\multicolumn{2}{c}{\sc MTM} \\
{} & {} & {} & 
\multicolumn{1}{c}{$\widehat{\theta}/\theta$} & 
\multicolumn{1}{c}{$\widehat{\sigma}/\sigma$} &
\multicolumn{1}{c}{$\widehat{\theta}/\theta$} & 
\multicolumn{1}{c|}{$\widehat{\sigma}/\sigma$} &
\multicolumn{1}{c}{$\widehat{\theta}/\theta$} & 
\multicolumn{1}{c}{$\widehat{\sigma}/\sigma$} &  
\multicolumn{1}{c}{$\widehat{\theta}/\theta$} & 
\multicolumn{1}{c|}{$\widehat{\sigma}/\sigma$} &
\multicolumn{1}{c}{$\widehat{\theta}/\theta$} & 
\multicolumn{1}{c}{$\widehat{\sigma}/\sigma$} & 
\multicolumn{1}{c}{$\widehat{\theta}/\theta$} & 
\multicolumn{1}{c|}{$\widehat{\sigma}/\sigma$} &
\multicolumn{1}{c}{$\widehat{\theta}/\theta$} & 
\multicolumn{1}{c}{$\widehat{\sigma}/\sigma$} & 
\multicolumn{1}{c}{$\widehat{\theta}/\theta$} & 
\multicolumn{1}{c}{$\widehat{\sigma}/\sigma$} \\
\hline\hline 
\multicolumn{1}{c|}{} &
\multicolumn{18}{c}{} \\[-2.50ex]
\multirow{16}{*}
{\rotatebox{90}{$u= 5.96 \times 10^3$}} &
\multicolumn{18}{c}
{Mean values of
	$\widehat{\theta}/\theta$ 
	and
	$\widehat{\sigma}/\sigma$.} \\
\cline{2-19} 
{} & 
\multicolumn{2}{c|}{MLE} & 
1.00 & 1.00 & 1.00 & 1.00 & 
1.00 & 1.00 & 1.00 & 1.00 & 
1.00 & 1.00 & 1.00 & 1.00 & 
1.00 & 1.00 & 1.00 & 1.00 \\
{} & 
0.05 & 0.05 & 
1.00 & 0.99 & 0.99 & 1.01 & 
1.00 & 1.00 & 1.00 & 1.00 & 
1.00 & 1.00 & 1.00 & 1.00 & 
1.00 & 1.00 & 1.00 & 1.00 \\
{} & 
0.10 & 0.10 & 
1.00 & 0.99 & 0.99 & 1.01 & 
1.00 & 1.00 & 1.00 & 1.00 & 
1.00 & 1.00 & 1.00 & 1.00 & 
1.00 & 1.00 & 1.00 & 1.00 \\
{} & 
0.15 & 0.15 & 
1.00 & 0.99 & 0.99 & 1.01 & 
1.00 & 1.00 & 1.00 & 1.00 & 
1.00 & 1.00 & 1.00 & 1.00 & 
1.00 & 1.00 & 1.00 & 1.00 \\
\cline{2-19} 
{} & 
0.00 & 0.05 & 
1.00 & 1.00 & 1.00 & 1.00 & 
1.00 & 1.00 & 1.00 & 1.00 & 
1.00 & 1.00 & 1.00 & 1.00 & 
1.00 & 1.00 & 1.00 & 1.00 \\
{} & 
0.00 & 0.10 & 
1.00 & 1.00 & 1.00 & 1.01 & 
1.00 & 1.00 & 1.00 & 1.00 & 
1.00 & 1.00 & 1.00 & 1.00 & 
1.00 & 1.00 & 1.00 & 1.00 \\
{} & 
0.00 & 0.25 & 
0.99 & 1.00 & 0.99 & 1.01 & 
1.00 & 1.00 & 1.00 & 1.00 & 
1.00 & 1.00 & 1.00 & 1.00 & 
1.00 & 1.00 & 1.00 & 1.00 \\
\cline{2-19} 
{} &
\multicolumn{18}{|c}
{Finite-sample efficiencies (RE) of MWMs and MTMs relative to MLEs.} \\
\cline{2-19} 
{} & 
\multicolumn{2}{c|}{MLE} & 
\multicolumn{2}{c}{0.97} & 
\multicolumn{2}{c}{0.98} &  
\multicolumn{2}{|c}{1.00} & 
\multicolumn{2}{c}{1.00} &  
\multicolumn{2}{|c}{1.00} & 
\multicolumn{2}{c}{1.00} & 
\multicolumn{2}{|c}{1.000} & 
\multicolumn{2}{c}{1.000} \\
{} & 
0.05 & 0.05 & 
\multicolumn{2}{c}{0.92} & 
\multicolumn{2}{c}{0.88} &  
\multicolumn{2}{|c}{0.95} & 
\multicolumn{2}{c}{0.92} &  
\multicolumn{2}{|c}{0.95} & 
\multicolumn{2}{c}{0.92} & 
\multicolumn{2}{|c}{0.945} & 
\multicolumn{2}{c}{0.913} \\
{} & 
0.10 & 0.10 & 
\multicolumn{2}{c}{0.84} & 
\multicolumn{2}{c}{0.77} &  
\multicolumn{2}{|c}{0.87} & 
\multicolumn{2}{c}{0.82} &  
\multicolumn{2}{|c}{0.88} & 
\multicolumn{2}{c}{0.83} & 
\multicolumn{2}{|c}{0.873} & 
\multicolumn{2}{c}{0.823} \\
{} & 
0.15 & 0.15 & 
\multicolumn{2}{c}{0.76} & 
\multicolumn{2}{c}{0.68} &  
\multicolumn{2}{|c}{0.79} & 
\multicolumn{2}{c}{0.73} &  
\multicolumn{2}{|c}{0.80} & 
\multicolumn{2}{c}{0.74} & 
\multicolumn{2}{|c}{0.796} & 
\multicolumn{2}{c}{0.734} \\
\cline{2-19} 
{} & 
0.00 & 0.05 & 
\multicolumn{2}{c}{0.92} & 
\multicolumn{2}{c}{0.89} &  
\multicolumn{2}{|c}{0.95} & 
\multicolumn{2}{c}{0.92} &  
\multicolumn{2}{|c}{0.95} & 
\multicolumn{2}{c}{0.92} & 
\multicolumn{2}{|c}{0.950} & 
\multicolumn{2}{c}{0.913} \\
{} & 
0.00 & 0.10 & 
\multicolumn{2}{c}{0.86} & 
\multicolumn{2}{c}{0.80} &  
\multicolumn{2}{|c}{0.89} & 
\multicolumn{2}{c}{0.84} &  
\multicolumn{2}{|c}{0.90} & 
\multicolumn{2}{c}{0.84} & 
\multicolumn{2}{|c}{0.892} & 
\multicolumn{2}{c}{0.841} \\
{} & 
0.00 & 0.25 & 
\multicolumn{2}{c}{0.68} & 
\multicolumn{2}{c}{0.59} &  
\multicolumn{2}{|c}{0.72} & 
\multicolumn{2}{c}{0.64} &  
\multicolumn{2}{|c}{0.73} & 
\multicolumn{2}{c}{0.65} & 
\multicolumn{2}{|c}{0.724} & 
\multicolumn{2}{c}{0.650} \\
\hline\hline  
\multicolumn{1}{c|}{} &
\multicolumn{18}{c}{} \\[-2.50ex]
\multirow{16}{*}
{\rotatebox{90}{$u= 1.54 \times 10^3$}} &
\multicolumn{18}{c}
{Mean values of
	$\widehat{\theta}/\theta$ 
	and
	$\widehat{\sigma}/\sigma$.} \\
\cline{2-19} 
{} & 
\multicolumn{2}{c|}{MLE} & 
1.00 & 1.00 & 1.00 & 1.00 & 
1.00 & 1.00 & 1.00 & 1.00 & 
1.00 & 1.00 & 1.00 & 1.00 & 
1.00 & 1.00 & 1.00 & 1.00 \\
{} & 
0.05 & 0.05 & 
1.00 & 0.98 & 0.99 & 1.00 & 
1.00 & 0.99 & 1.00 & 1.00 & 
1.00 & 1.00 & 1.00 & 1.00 & 
1.00 & 1.00 & 1.00 & 1.00 \\
{} & 
0.10 & 0.10 & 
1.00 & 0.99 & 0.99 & 1.01 & 
1.00 & 1.00 & 1.00 & 1.00 & 
1.00 & 1.00 & 1.00 & 1.00 & 
1.00 & 1.00 & 1.00 & 1.00 \\
{} & 
0.15 & 0.15 & 
1.00 & 0.99 & 0.99 & 1.01 & 
1.00 & 1.00 & 1.00 & 1.00 & 
1.00 & 1.00 & 1.00 & 1.00 & 
1.00 & 1.00 & 1.00 & 1.00 \\
\cline{2-19} 
{} & 
0.00 & 0.05 & 
1.00 & 0.98 & 1.00 & 1.00 & 
1.00 & 1.00 & 1.00 & 1.00 & 
1.00 & 1.00 & 1.00 & 1.00 & 
1.00 & 1.00 & 1.00 & 1.00 \\
{} & 
0.00 & 0.10 & 
1.00 & 1.00 & 0.99 & 1.01 & 
1.00 & 1.00 & 1.00 & 1.00 & 
1.00 & 1.00 & 1.00 & 1.00 & 
1.00 & 1.00 & 1.00 & 1.00 \\
{} & 
0.00 & 0.25 & 
0.99 & 1.00 & 0.99 & 1.01 & 
1.00 & 1.00 & 1.00 & 1.00 & 
1.00 & 1.00 & 1.00 & 1.00 & 
1.00 & 1.00 & 1.00 & 1.00 \\
\cline{2-19} 
{} &
\multicolumn{18}{|c}
{Finite-sample efficiencies (RE) of MWMs and MTMs relative to MLEs.} \\
\cline{2-19} 
{} & 
\multicolumn{2}{c|}{MLE} & 
\multicolumn{2}{c}{0.95} & 
\multicolumn{2}{c}{0.96} &  
\multicolumn{2}{|c}{0.98} & 
\multicolumn{2}{c}{0.98} &  
\multicolumn{2}{|c}{1.00} & 
\multicolumn{2}{c}{1.00} & 
\multicolumn{2}{|c}{1.000} & 
\multicolumn{2}{c}{1.000} \\
{} & 
0.05 & 0.05 & 
\multicolumn{2}{c}{1.10} & 
\multicolumn{2}{c}{0.97} &  
\multicolumn{2}{|c}{1.10} & 
\multicolumn{2}{c}{0.96} &  
\multicolumn{2}{|c}{1.10} & 
\multicolumn{2}{c}{0.98} & 
\multicolumn{2}{|c}{0.994} & 
\multicolumn{2}{c}{0.960} \\
{} & 
0.10 & 0.10 & 
\multicolumn{2}{c}{0.89} & 
\multicolumn{2}{c}{0.81} &  
\multicolumn{2}{|c}{0.90} & 
\multicolumn{2}{c}{0.84} &  
\multicolumn{2}{|c}{0.91} & 
\multicolumn{2}{c}{0.86} & 
\multicolumn{2}{|c}{0.919} & 
\multicolumn{2}{c}{0.865} \\
{} & 
0.15 & 0.15 & 
\multicolumn{2}{c}{0.80} & 
\multicolumn{2}{c}{0.71} &  
\multicolumn{2}{|c}{0.82} & 
\multicolumn{2}{c}{0.75} &  
\multicolumn{2}{|c}{0.83} & 
\multicolumn{2}{c}{0.77} & 
\multicolumn{2}{|c}{0.837} & 
\multicolumn{2}{c}{0.772} \\
\cline{2-19} 
{} & 
0.00 & 0.05 & 
\multicolumn{2}{c}{1.11} & 
\multicolumn{2}{c}{0.97} &  
\multicolumn{2}{|c}{1.11} & 
\multicolumn{2}{c}{0.97} &  
\multicolumn{2}{|c}{1.11} & 
\multicolumn{2}{c}{0.98} & 
\multicolumn{2}{|c}{1.000} & 
\multicolumn{2}{c}{0.964} \\
{} & 
0.00 & 0.10 & 
\multicolumn{2}{c}{0.91} & 
\multicolumn{2}{c}{0.84} &  
\multicolumn{2}{|c}{0.92} & 
\multicolumn{2}{c}{0.86} &  
\multicolumn{2}{|c}{0.93} & 
\multicolumn{2}{c}{0.88} & 
\multicolumn{2}{|c}{0.938} & 
\multicolumn{2}{c}{0.884} \\
{} & 
0.00 & 0.25 & 
\multicolumn{2}{c}{0.72} & 
\multicolumn{2}{c}{0.61} &  
\multicolumn{2}{|c}{0.75} & 
\multicolumn{2}{c}{0.66} &  
\multicolumn{2}{|c}{0.76} & 
\multicolumn{2}{c}{0.67} & 
\multicolumn{2}{|c}{0.762} & 
\multicolumn{2}{c}{0.684} \\
\hline\hline  
\end{tabular} \\[5pt]
{
\small 
{\sc Note:}
The standard errors for the entire entries 
in this table are reported to be $\le 0.0015$.
}
}
\end{table}

\subsection{Convergence and Relative Bias}

\begin{table}[hbt!]
\caption{
Lognormal payment-per-loss 
actuarial loss scenario,
$LN(w_{0}=1,\theta=4,\sigma=2)$ with $d = 3$
and two selected values of right-censoring point $u$.
The entries are mean values 
based on 10,000 samples.
}
\label{table:PPZ_SimStudy_TW}
\centering
{\scriptsize 
\begin{tabular}{c|cc|cccc|cccc|cccc|cccc}
\cline{2-19}
{} & 
\multicolumn{2}{c|}{Proportion} & 
\multicolumn{4}{|c|}{$n=100$} & 
\multicolumn{4}{|c|}{$n=500$} &
\multicolumn{4}{|c|}{$n=1000$} &
\multicolumn{4}{|c}{$n \to \infty$} \\
\cline{2-19} 
& & & & & & & & & 
& & & & & & & & & \\[-2.50ex]
{} & 
\multirow{2}{*}{$a$} & 
\multirow{2}{*}{$b$} & 
\multicolumn{2}{|c}{\sc MWM} & 
\multicolumn{2}{c|}{\sc MTM} & 
\multicolumn{2}{c}{\sc MWM} & 
\multicolumn{2}{c|}{\sc MTM} &
\multicolumn{2}{c}{\sc MWM} & 
\multicolumn{2}{c|}{\sc MTM} &
\multicolumn{2}{c}{\sc MWM} & 
\multicolumn{2}{c}{\sc MTM} \\
{} & {} & {} & 
\multicolumn{1}{c}{$\widehat{\theta}/\theta$} & 
\multicolumn{1}{c}{$\widehat{\sigma}/\sigma$} &
\multicolumn{1}{c}{$\widehat{\theta}/\theta$} & 
\multicolumn{1}{c|}{$\widehat{\sigma}/\sigma$} &
\multicolumn{1}{c}{$\widehat{\theta}/\theta$} & 
\multicolumn{1}{c}{$\widehat{\sigma}/\sigma$} &  
\multicolumn{1}{c}{$\widehat{\theta}/\theta$} & 
\multicolumn{1}{c|}{$\widehat{\sigma}/\sigma$} &
\multicolumn{1}{c}{$\widehat{\theta}/\theta$} & 
\multicolumn{1}{c}{$\widehat{\sigma}/\sigma$} & 
\multicolumn{1}{c}{$\widehat{\theta}/\theta$} & 
\multicolumn{1}{c|}{$\widehat{\sigma}/\sigma$} &
\multicolumn{1}{c}{$\widehat{\theta}/\theta$} & 
\multicolumn{1}{c}{$\widehat{\sigma}/\sigma$} & 
\multicolumn{1}{c}{$\widehat{\theta}/\theta$} & 
\multicolumn{1}{c}{$\widehat{\sigma}/\sigma$} \\
\hline\hline 
\multicolumn{1}{c|}{} &
\multicolumn{18}{c}{} \\[-2.50ex]
\multirow{16}{*}
{\rotatebox{90}{$u= 5.96 \times 10^3$}} &
\multicolumn{18}{c}
{Mean values of
	$\widehat{\theta}/\theta$ 
	and
	$\widehat{\sigma}/\sigma$.} \\
\cline{2-19} 
{} & 
\multicolumn{2}{c|}{MLE} & 
1.00 & 0.99 & 1.00 & 0.99 & 
1.00 & 1.00 & 1.00 & 1.00 & 
1.00 & 1.00 & 1.00 & 1.00 & 
1.00 & 1.00 & 1.00 & 1.00 \\
{} & 
0.10 & 0.10 & 
1.00 & 0.99 & 1.00 & 1.00 & 
1.00 & 1.00 & 1.00 & 1.00 & 
1.00 & 1.00 & 1.00 & 1.00 & 
1.00 & 1.00 & 1.00 & 1.00 \\
{} & 
0.15 & 0.15 & 
1.00 & 0.99 & 1.00 & 1.00 & 
1.00 & 1.00 & 1.00 & 1.00 & 
1.00 & 1.00 & 1.00 & 1.00 & 
1.00 & 1.00 & 1.00 & 1.00 \\
{} & 
0.25 & 0.25 & 
1.00 & 0.98 & 1.00 & 1.01 & 
1.00 & 1.00 & 1.00 & 1.00 & 
1.00 & 1.00 & 1.00 & 1.00 & 
1.00 & 1.00 & 1.00 & 1.00 \\
\cline{2-19} 
{} & 
0.10 & 0.05 & 
1.00 & 0.99 & 1.00 & 1.00 & 
1.00 & 1.00 & 1.00 & 1.00 & 
1.00 & 1.00 & 1.00 & 1.00 & 
1.00 & 1.00 & 1.00 & 1.00 \\
{} & 
0.10 & 0.15 & 
1.00 & 0.99 & 1.00 & 1.00 & 
1.00 & 1.00 & 1.00 & 1.00 & 
1.00 & 1.00 & 1.00 & 1.00 & 
1.00 & 1.00 & 1.00 & 1.00 \\
{} & 
0.10 & 0.25 & 
1.00 & 0.99 & 1.00 & 1.00 & 
1.00 & 1.00 & 1.00 & 1.00 & 
1.00 & 1.00 & 1.00 & 1.00 & 
1.00 & 1.00 & 1.00 & 1.00 \\
\cline{2-19} 
{} &
\multicolumn{18}{|c}
{Finite-sample efficiencies (RE) of MWMs and MTMs relative to MLEs.} \\
\cline{2-19} 
{} & 
\multicolumn{2}{|c|}{MLE} &
\multicolumn{2}{c}{0.99} & 
\multicolumn{2}{c}{1.00} &  
\multicolumn{2}{|c}{1.00} & 
\multicolumn{2}{c}{1.00} &  
\multicolumn{2}{|c}{1.00} & 
\multicolumn{2}{c}{1.00} & 
\multicolumn{2}{|c}{1.000} & 
\multicolumn{2}{c}{1.000} \\
{} & 
0.10 & 0.10 & 
\multicolumn{2}{c}{0.87} & 
\multicolumn{2}{c}{0.81} &  
\multicolumn{2}{|c}{0.88} & 
\multicolumn{2}{c}{0.82} &  
\multicolumn{2}{|c}{0.88} & 
\multicolumn{2}{c}{0.79} & 
\multicolumn{2}{|c}{0.891} & 
\multicolumn{2}{c}{0.810} \\
{} & 
0.15 & 0.15 & 
\multicolumn{2}{c}{0.78} & 
\multicolumn{2}{c}{0.71} &  
\multicolumn{2}{|c}{0.80} & 
\multicolumn{2}{c}{0.72} &  
\multicolumn{2}{|c}{0.79} & 
\multicolumn{2}{c}{0.70} & 
\multicolumn{2}{|c}{0.791} & 
\multicolumn{2}{c}{0.712} \\
{} & 
0.25 & 0.25 & 
\multicolumn{2}{c}{0.60} & 
\multicolumn{2}{c}{0.53} &  
\multicolumn{2}{|c}{0.61} & 
\multicolumn{2}{c}{0.54} &  
\multicolumn{2}{|c}{0.61} & 
\multicolumn{2}{c}{0.52} & 
\multicolumn{2}{|c}{0.603} & 
\multicolumn{2}{c}{0.534} \\
\cline{2-19} 
{} & 
0.10 & 0.05 & 
\multicolumn{2}{c}{0.92} & 
\multicolumn{2}{c}{0.87} &  
\multicolumn{2}{|c}{0.93} & 
\multicolumn{2}{c}{0.87} &  
\multicolumn{2}{|c}{0.93} & 
\multicolumn{2}{c}{0.85} & 
\multicolumn{2}{|c}{0.927} & 
\multicolumn{2}{c}{0.863} \\
{} & 
0.10 & 0.15 & 
\multicolumn{2}{c}{0.83} & 
\multicolumn{2}{c}{0.76} &  
\multicolumn{2}{|c}{0.84} & 
\multicolumn{2}{c}{0.77} &  
\multicolumn{2}{|c}{0.84} & 
\multicolumn{2}{c}{0.75} & 
\multicolumn{2}{|c}{0.854} & 
\multicolumn{2}{c}{0.761} \\
{} & 
0.10 & 0.25 & 
\multicolumn{2}{c}{0.74} & 
\multicolumn{2}{c}{0.66} &  
\multicolumn{2}{|c}{0.75} & 
\multicolumn{2}{c}{0.68} &  
\multicolumn{2}{|c}{0.75} & 
\multicolumn{2}{c}{0.66} & 
\multicolumn{2}{|c}{0.776} & 
\multicolumn{2}{c}{0.667} \\
\hline\hline  
\multicolumn{1}{c|}{} &
\multicolumn{18}{c}{} \\[-2.50ex]
\multirow{16}{*}
{\rotatebox{90}{$u= 1.54 \times 10^3$}} &
\multicolumn{18}{c}
{Mean values of
	$\widehat{\theta}/\theta$ 
	and
	$\widehat{\sigma}/\sigma$.} \\
\cline{2-19} 
{} & 
\multicolumn{2}{c|}{MLE} & 
1.00 & 1.00 & 1.00 & 1.00 & 
1.00 & 1.00 & 1.00 & 1.00 & 
1.00 & 1.00 & 1.00 & 1.00 & 
1.00 & 1.00 & 1.00 & 1.00 \\
{} & 
0.10 & 0.10 & 
1.00 & 0.99 & 1.00 & 1.00 & 
1.00 & 1.00 & 1.00 & 1.00 & 
1.00 & 1.00 & 1.00 & 1.00 & 
1.00 & 1.00 & 1.00 & 1.00 \\
{} & 
0.15 & 0.15 & 
1.00 & 0.99 & 1.00 & 1.00 & 
1.00 & 1.00 & 1.00 & 1.00 & 
1.00 & 1.00 & 1.00 & 1.00 & 
1.00 & 1.00 & 1.00 & 1.00 \\
{} & 
0.25 & 0.25 & 
1.00 & 0.98 & 1.00 & 1.01 & 
1.00 & 1.00 & 1.00 & 1.00 & 
1.00 & 1.00 & 1.00 & 1.00 & 
1.00 & 1.00 & 1.00 & 1.00 \\
\cline{2-19} 
{} & 
0.10 & 0.05 & 
1.00 & 0.98 & 1.00 & 1.00 & 
1.00 & 1.00 & 1.00 & 1.00 & 
1.00 & 1.00 & 1.00 & 1.00 & 
1.00 & 1.00 & 1.00 & 1.00 \\
{} & 
0.10 & 0.15 & 
1.00 & 0.99 & 1.00 & 1.00 & 
1.00 & 1.00 & 1.00 & 1.00 & 
1.00 & 1.00 & 1.00 & 1.00 & 
1.00 & 1.00 & 1.00 & 1.00 \\
{} & 
0.10 & 0.25 & 
1.00 & 0.99 & 1.00 & 1.00 & 
1.00 & 1.00 & 1.00 & 1.00 & 
1.00 & 1.00 & 1.00 & 1.00 & 
1.00 & 1.00 & 1.00 & 1.00 \\
\cline{2-19} 
{} &
\multicolumn{18}{|c}
{Finite-sample efficiencies (RE) of MWMs and MTMs relative to MLEs.} \\
\cline{2-19} 
{} & 
\multicolumn{2}{|c|}{MLE} &
\multicolumn{2}{c}{1.00} & 
\multicolumn{2}{c}{0.97} &  
\multicolumn{2}{|c}{1.02} & 
\multicolumn{2}{c}{1.02} &  
\multicolumn{2}{|c}{1.03} & 
\multicolumn{2}{c}{1.03} & 
\multicolumn{2}{|c}{1.000} & 
\multicolumn{2}{c}{1.000} \\
{} & 
0.10 & 0.10 & 
\multicolumn{2}{c}{0.92} & 
\multicolumn{2}{c}{0.83} &  
\multicolumn{2}{|c}{0.92} & 
\multicolumn{2}{c}{0.82} &  
\multicolumn{2}{|c}{0.92} & 
\multicolumn{2}{c}{0.83} & 
\multicolumn{2}{|c}{0.923} & \multicolumn{2}{c}{0.839} \\
{} & 
0.15 & 0.15 & 
\multicolumn{2}{c}{0.82} & 
\multicolumn{2}{c}{0.73} &  
\multicolumn{2}{|c}{0.82} & 
\multicolumn{2}{c}{0.72} &  
\multicolumn{2}{|c}{0.83} & 
\multicolumn{2}{c}{0.74} & 
\multicolumn{2}{|c}{0.819} & \multicolumn{2}{c}{0.738} \\
{} & 
0.25 & 0.25 & 
\multicolumn{2}{c}{0.63} & 
\multicolumn{2}{c}{0.54} &  
\multicolumn{2}{|c}{0.63} & 
\multicolumn{2}{c}{0.54} &  
\multicolumn{2}{|c}{0.63} & 
\multicolumn{2}{c}{0.56} & 
\multicolumn{2}{|c}{0.625} & 
\multicolumn{2}{c}{0.553} \\
\cline{2-19} 
{} & 
0.10 & 0.05 & 
\multicolumn{2}{c}{1.05} & 
\multicolumn{2}{c}{0.91} &  
\multicolumn{2}{|c}{1.05} & 
\multicolumn{2}{c}{0.89} &  
\multicolumn{2}{|c}{1.04} & 
\multicolumn{2}{c}{0.90} & 
\multicolumn{2}{|c}{0.961} & 
\multicolumn{2}{c}{0.894} \\
{} & 
0.10 & 0.15 & 
\multicolumn{2}{c}{0.87} & 
\multicolumn{2}{c}{0.77} &  
\multicolumn{2}{|c}{0.87} & 
\multicolumn{2}{c}{0.77} &  
\multicolumn{2}{|c}{0.87} & 
\multicolumn{2}{c}{0.79} & 
\multicolumn{2}{|c}{0.884} & 
\multicolumn{2}{c}{0.788} \\
{} & 
0.10 & 0.25 & 
\multicolumn{2}{c}{0.77} & 
\multicolumn{2}{c}{0.68} &  
\multicolumn{2}{|c}{0.77} & 
\multicolumn{2}{c}{0.68} &  
\multicolumn{2}{|c}{0.78} & 
\multicolumn{2}{c}{0.69} & 
\multicolumn{2}{|c}{0.804} & 
\multicolumn{2}{c}{0.690} \\
\hline\hline  
\end{tabular} \\[5pt]
{
\small 
{\sc Note:}
The standard errors for the entire entries 
in this table are reported to be $\le 0.0013$.
}
}
\end{table}

Simulation results of both MTM and MWM are recorded in Tables 
\ref{table:PPY_SimStudy_TW} (for payment-per-payment variable $Y$) 
and  \ref{table:PPZ_SimStudy_TW} (for payment-per-payment variable $Z$). 
The entries of the last column ($n\to \infty$) represent analytic results that 
are included in Section 5. 
It helps us compare large-sample properties with 
small-sample properties when the deductible and policy limit occur.

The relative bias is defined as the ratio
of the expectation of parameter estimate to its true value, 
thus making the value of one
the target. As is seen in Table \ref{table:PPY_SimStudy_TW}, 
all MWM and MTM estimators successfully estimate 
both the log-location parameter $\mu$ and scale parameter $\sigma$.
Indeed, they practically become unbiased for samples of size $n \geq 500$. 
The situation is similar in Table \ref{table:PPZ_SimStudy_TW}. 
Furthermore, we notice that
the simulated RE’s of these estimators for $n \geq 500$ are almost identical
to the corresponding ARE’s for large policy limit $u=5.96\times 10^3$. 
When $u$ moves to $1.54\times 10^3$, however, the RE's of MWM in
finite size samples are far from corresponding ARE for $b=0.05$ 
and this is because the condition \eqref{eqn:CondsStarB}
is violated which leads to poor parameter estimation.

\subsection{Risk Measure Sensitivity}

In this section, we study sensitivity of our estimators to an outlier 
$\mathcal{O}$ that is placed at various locations within the sample 
ranging from $d$ to $u$. The outlier can be viewed as a new observation 
for the original sample, thus when all the estimates are recomputed,
they can be plotted against the location of $\mathcal{O}$ resulting 
in sensitivity curves. 
To see what impact the outlier has on the riskiness of the ground-up 
variable, $W \sim LN(w_{0},\theta,\sigma)$ as defined in 
Section \ref{sec:PP_DataScenarios1}, we consider four risk measures: 
Mean 
$(\mu$, $\mu_{\mbox{\tiny MTM}}$, and $\mu_{\mbox{\tiny MWM}})$,
Value at Risk $(\mbox{VaR}_{p})$,
Tail Value at Risk $(\mbox{VTaR}_{p})$, 
and Proportional Hazard Distortion Risk Measure
$(\mbox{PHDRM}_{p})$. Each of them is a special case of the general 
distortion risk measure
\begin{align}
\label{eqn:RiskMeasureDef1}
R(F_{W}) 
& = 
\int_{0}^{1} 
F_{W}^{-1}(v)J(v) \, dv
+ 
\sum_{j=1}^{m}\beta_{j} F_{W}^{-1}(p_{j}),
\end{align}
where $F_{W}$ is the cdf of $W$.
More details are referred in \cite{MR595165}, p. 265 and 
\cite{MR3765327}. 
The specific  components of each risk measure given by 
equation \eqref{eqn:RiskMeasureDef1} are listed in 
Table \ref{table:Measure}.

\begin{table}[htb]
\centering
\caption{Mean,
Value at Risk $(\mbox{VaR}_{p})$,
Tail Value at Risk $(\mbox{VTaR}_{p})$,
and Proportional Hazard Distortion Risk Measure
$(\mbox{PHDRM}_{p})$ in general risk measure format.}
\label{table:Measure}
{\small
\centering
\begin{tabular}{r|lccccl}
\cline{2-7}
{} & 
Risk Measure & 
$J(v)$ & $m$ & 
$(\beta_{1},\beta_{2})$ & 
$(p_{1},p_{2})$ & $R(F_{W})$ \\
\hline\hline
\multirow{8}{*}{\rotatebox{90}{\sc Mean}} &
$\mu$ & 1 & 0 & - & - & 
$w_{0} + e^{\theta+\frac{1}{2}\sigma^2}$ \\
\cline{2-7}
& & & & & & \\[-2.50ex]
{} & 
\multirow{2}{*}{$\mu_{\mbox{\tiny MTM}}$} & 
\multirow{2}{*}{$\dfrac{1}{\tau} \ID\{ a \le v \le \bar{b} \}$} & 
\multirow{2}{*}{0} &
\multirow{2}{*}{-} &
\multirow{2}{*}{-} & $w_{0} 
+ 
\tau^{-1} 
e^{\theta+\frac{1}{2}\sigma^2}\cdot$ \\
& & & & & & 
$\hspace{0.15in}
\left( 
\Phi
\left(
\Phi^{-1}(\bar{b})-\sigma
\right) 
-\Phi\big(\Phi^{-1}(a)-\sigma\big)
\right)$ \\
\cline{2-7}
& & & & & & \\[-2.50ex]
{} & 
\multirow{2}{*}{$\mu_{\mbox{\tiny MWM}}$} &
\multirow{2}{*}{$\ID\{ a \le v \le \bar{b} \}$} &
\multirow{2}{*}{2} &
\multirow{2}{*}{$(a,b)$}&
\multirow{2}{*}{$(a,\bar{b})$} & $w_{0}
+ 
e^{\theta} \cdot
\bigg[ 
a 
e^{\sigma \Phi^{-1}(a)}
+
b
e^{\sigma \Phi^{-1}(\bar{b})}+ e^{\frac{1}{2}\sigma^2} \cdot
$ \\
& & & & & & $\hspace{0.15in}
\left(
\Phi
\left(
\Phi^{-1}(\bar{b})-\sigma
\right)
-\Phi
\left(
\Phi^{-1}(a)-\sigma
\right)
\right)\bigg]$ \\[1.2ex]
\hline
\multirow{6}{*}{\rotatebox{90}{\sc Other}} & 
$\mbox{VaR}_{p}$ & 0 & 1 & - &
$p_{1} 
= 
p \in [0,1)$&
$w_{0} + e^{\theta + \sigma \Phi^{-1}(p)}
$ \\[1.2ex]
\cline{2-7}
& & & & & & \\[-2.50ex]
{} & $\mbox{TVaR}_{p}$ & $\dfrac{1}{\bar{p}}
\ID\{ p \le v \le 1 \}$ & 0 & - & - &
$w_{0} 
+\dfrac{e^{\theta+\frac{1}{2}\sigma^2}}
{\bar{p}}
\Phi 
\left( 
\sigma - \Phi^{-1}(p)
\right)
$ \\[1.2ex]
\cline{2-7}
{} & 
\multirow{2}{*}{$\mbox{PHDRM}_{p}$} &
\multirow{2}{*}{$p(1-v)^{p-1}$} & 
\multirow{2}{*}{0} & \multirow{2}{*}{-} &
\multirow{2}{*}{-} &
\multirow{2}{*}{$\displaystyle{\int_{0}^{\infty}
\left[
1 - 
\Phi 
\left(
\frac{\ln{w} 
-
\theta}
{\sigma}
\right)
\right]^p \, dw}
$} \\
{} & & & & & & \\[1.2ex]
\hline
\end{tabular}
\centering
{\sc Note}: When $m=0, (\beta_{1},\beta_{2})$ and
$(p_{1},p_{2})$ are not available and the empty spaces 
are filled with $``-''$.
}
\end{table}

\subsubsection{Payments {\em Y}} 

\begin{figure}[hbt!]
\centering
\includegraphics[width=0.95\textwidth] 
{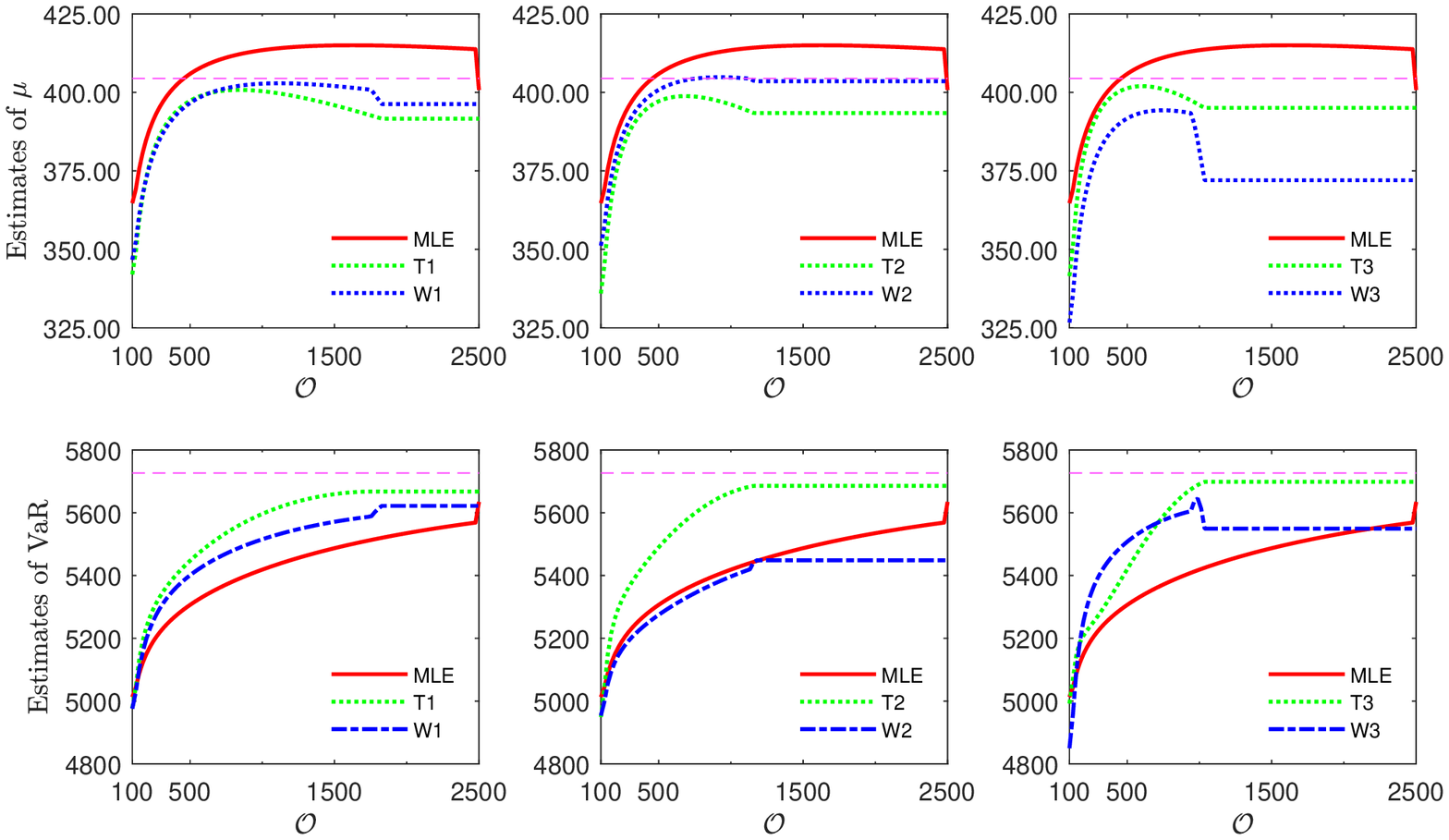}
\includegraphics[width=0.95\textwidth] 
{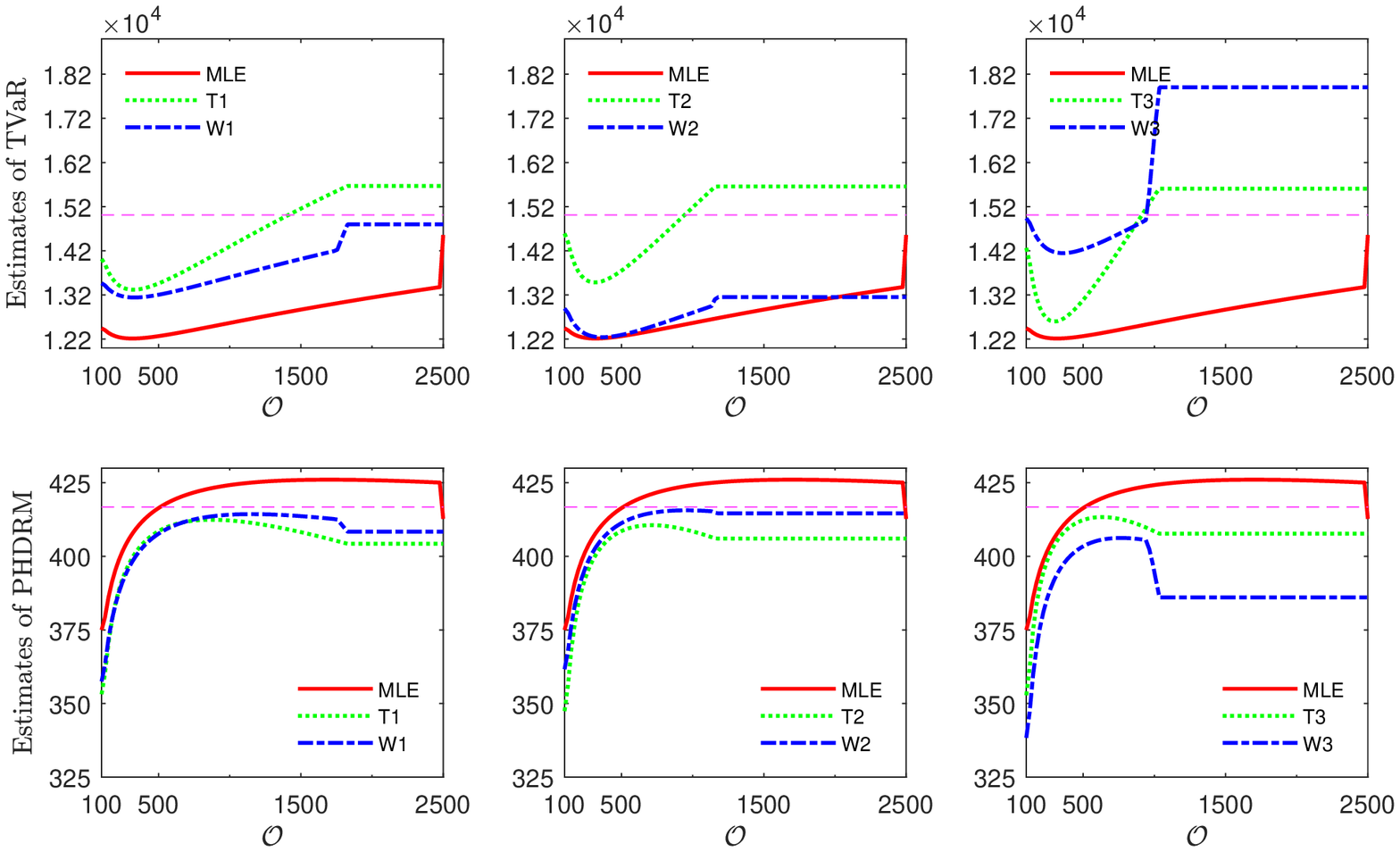}
\caption{
Sensitivity curves of MLE (solid curve), 
trimmed (dotted curve), and winsorized (dash-dot curve)
estimators of
$\mu = 404.43$, 
$\mbox{VaR}_{0.99} = 5,726.56$, 
$\mbox{TVaR}_{0.99} = 15,011.80$, and 
$\mbox{PHDRM}_{0.99} = 416.74$
as the outlier $\mathcal{O}$ 
increases from $100$ to $2,500$ under the {\em payment-per-payment\/} 
data scenario. 
In each plot, the theoretical target is marked as dashed line.
}
\label{fig:PPY_Graphs_RNG_83_PiP_99}
\end{figure}

A sample of size 100 generated from 
$W \sim LN(w_0=1,\theta=4, \sigma=2)$
with $d = 100$ (about 62\% left truncation),
$u =  2.5 \times 10^{3}$ (about 3\% right censoring) 
is given in Appendix \ref{sec:AppendixB}, 
Table \ref{tab:Sensitivity_PPY}.
The parameters $\theta$ and $\sigma$ are first estimated
via MLE, 
$T_{1} \equiv MTM(a=0.00, b=0.10)$,
$T_{2} \equiv MTM(a=0.00, b=0.15)$,
$T_{3} \equiv MTM(a=0.00, b=0.20)$,
$W_{1} \equiv MWM(a=0.00, b=0.10)$,
$W_{2} \equiv MWM(a=0.00, b=0.15)$,
$W_{3} \equiv MWM(a=0.00, b=0.20)$,
by placing an outlier $\mathcal{O}$ 
at various locations between 
$d = 100$ and $u = 2500$ 
(or, in terms of $Y$,
between $0$ and $2400$).
The right trimming/winsorizing 
proportion $a$ and $b$ is chosen to satisfy 
the condition $1-b \le s^{*} = 0.93$.
Since $F_{W}(d)$ is already approximately 62\%,
the left trimming/winsorizing proportion $a$
is fixed as $0$. 
The corresponding risk measure sensitivity 
curves are then reported in Figure 
\ref{fig:PPY_Graphs_RNG_83_PiP_99}.

The stability exhibited by the curves of $T3-$ and 
$W3-$ estimators is about 63\% of possible data range 
($63\% \approx (2500-1000)/(2500-100)$), and this width 
narrows down to about 33\%  for $T_{1}, W_{1}$ and 
to about 58\% for $T_{2}, W_{2}$ when the proportion $b$ 
tends to be small. 
Besides, all MTM and MWM risks measures 
of $\mu, \text{VaR}_{0.99}, \text{TVaR}_{0.99}$ and 
$\text{PHDRM}_{0.99}$ remain stable when the new 
observation $\mathcal{O}$ occurs near the upper 
boundary of 2500, whereas the equivalent estimates 
based on MLE show a sudden jump/drop at that 
right-censoring threshold. 

As seen in the sensitivity curve, MTM and MWM based 
$\text{VaR}_{0.99}$ and $\text{TVaR}_{0.99}$ values 
are much closer to the theoretical targets (marked 
as dashed lines in Figure \ref{fig:PPY_Graphs_RNG_83_PiP_99}) 
than those of MLE results at most locations. This indicates 
that without the more extreme observations, MTM and MWM 
estimators can still capture the behavior of shortfall 
above the 99\% threshold and generate adequate risk 
control measures.

\subsubsection{Payments {\em Z}} 

\begin{figure}[htb!]
\centering
\includegraphics[width=0.80\textwidth] 
{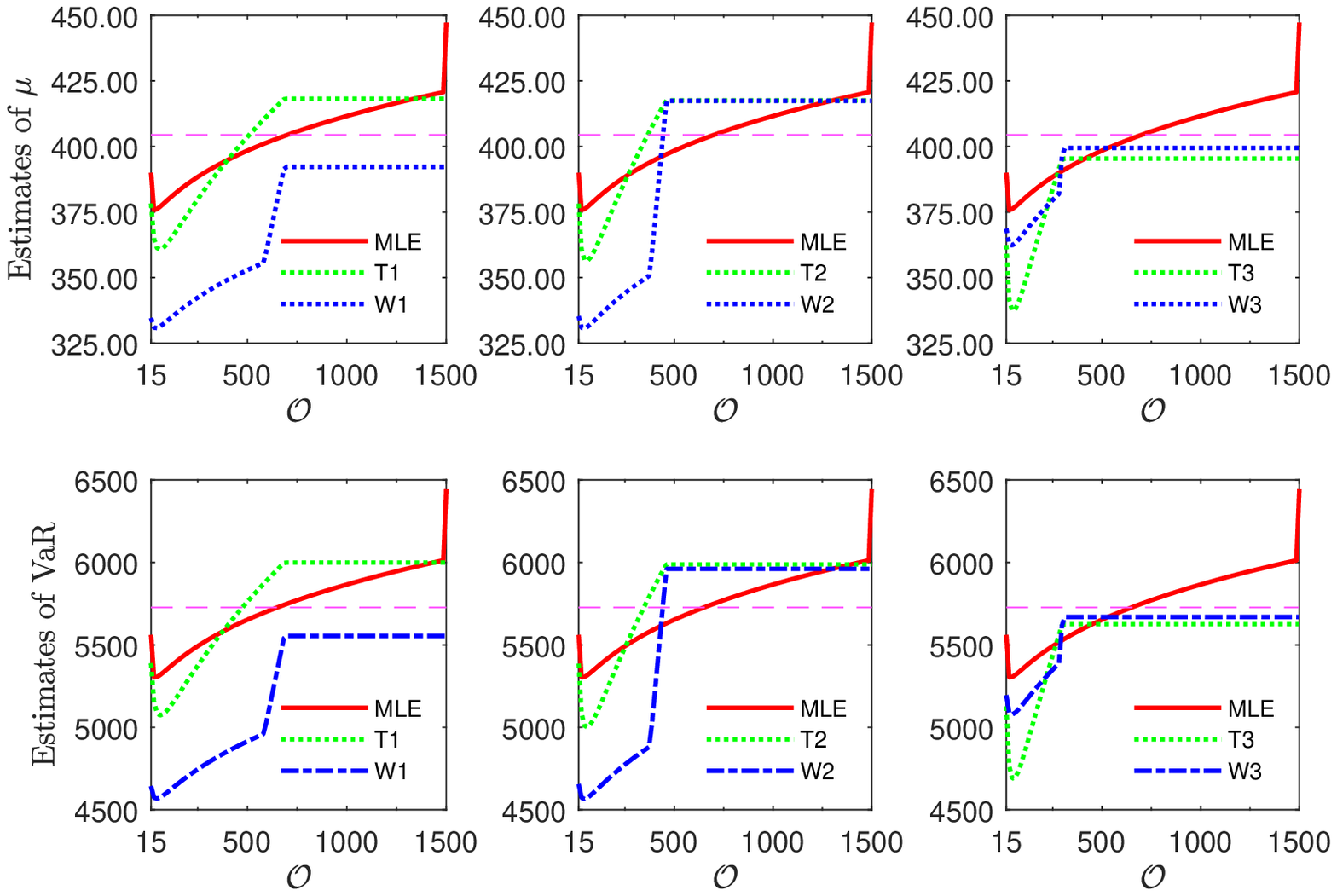}
\includegraphics[width=0.80\textwidth] 
{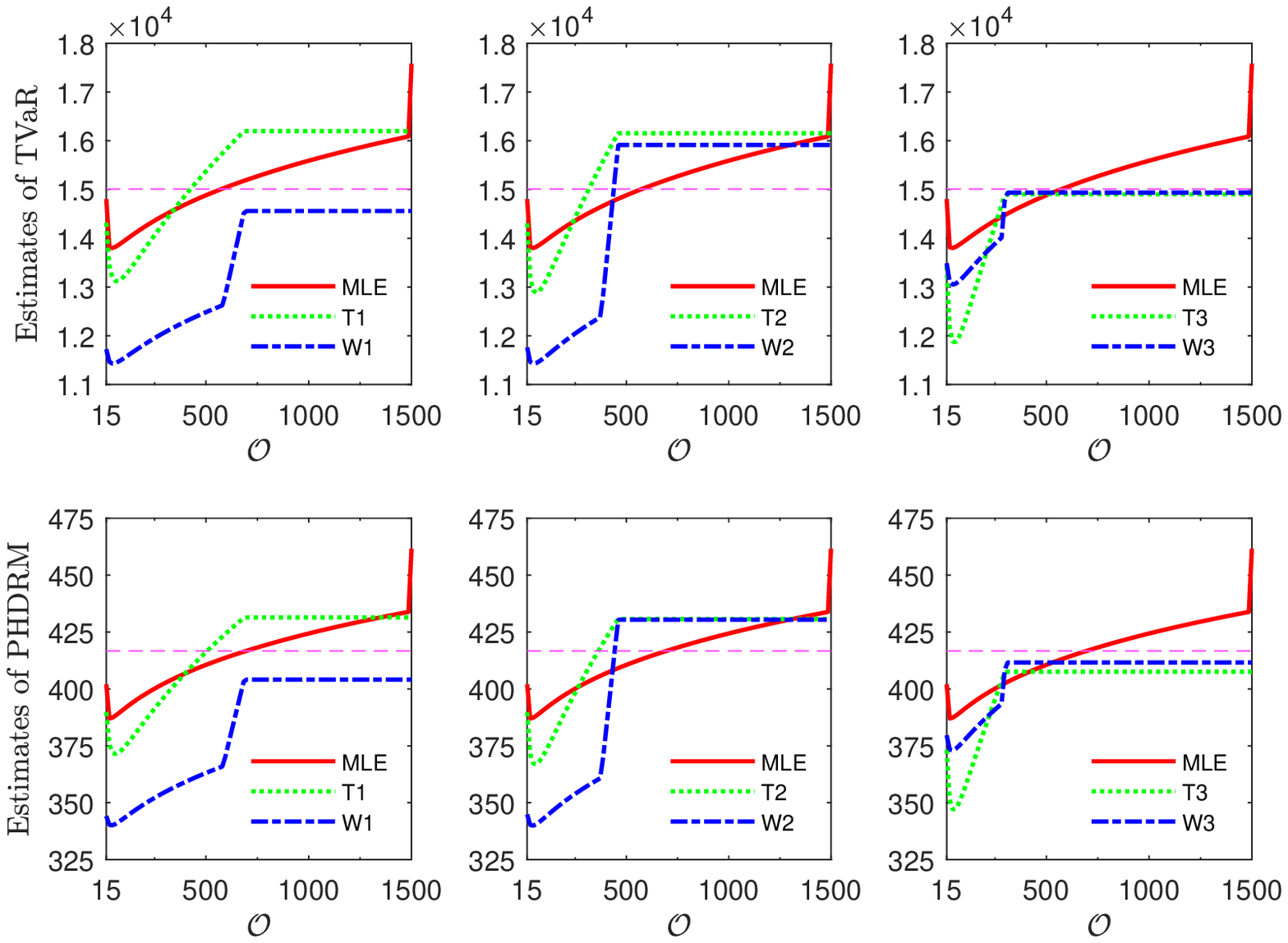}
\caption{
Sensitivity curves of MLE (solid curve), 
trimmed (dotted curve), and winsorized (dash-dot curve)
estimators of
$\mu = 404.43$, 
$\mbox{VaR}_{0.99} = 5,726.56$, 
$\mbox{TVaR}_{0.99} = 15,011.80$, and 
$\mbox{PHDRM}_{0.99} = 416.74$
as the outlier $\mathcal{O}$ 
increases from $15$ to $1,500$ under the {\em payment-per-loss\/} 
data scenario. 
In each plot, the theoretical target is marked as dashed line.
}
\label{fig:PPZ_Graphs_RNG_35_PiP_99}
\end{figure}

A sample of size 100 generated from 
$W \sim LN(w_0=1,\theta=4, \sigma=2)$
with $d = 15$ (about 25\% left truncation), 
$u =  1.5 \times 10^{3}$ (about 5\% right censoring)
is given in Appendix \ref{sec:AppendixB},
Table \ref{tab:Sensitivity_PPZ}.
Again, 
the parameters $\theta$ and $\sigma$ 
are first estimated
via MLE, 
$T_{1} \equiv MTM(a=0.25, b=0.10)$,
$T_{2} \equiv MTM(a=0.25, b=0.15)$,
$T_{3} \equiv MTM(a=0.25, b=0.20)$,
$W_{1} \equiv MWM(a=0.25, b=0.10)$,
$W_{2} \equiv MWM(a=0.25, b=0.15)$,
$W_{3} \equiv MWM(a=0.25, b=0.20)$,
by placing an outlier $\mathcal{O}$ 
at various locations between 
$d = 15$ and $u = 1500$ 
(or, in terms of $Z$,
between $0$ and 1,485).
The corresponding risk measure sensitivity 
curves are then reported in Figure 
\ref{fig:PPZ_Graphs_RNG_35_PiP_99}.
As mentioned in Section \ref{sec:PPZ_MWM}, 
the trimming/winsorizing proportion $a$ and $b$ are 
chosen to satisfy the condition
$
0 \le F_{W}(d) \le a < 1-b \le F_{W}(u) \le 1.
$

It is obvious that the pattern of the four risk measures 
follows the same direction. The MLE based sensitivity 
curves are influenced by all locations of $\mathcal{O}$; 
in particular, a larger jump near the upper boundary of 1500. 
This indicates that even an extreme claim has been censored 
to the threshold, it still highly affects the estimation 
procedure of the maximum likelihood method, leading to an 
over- or under-estimate of risk pricing in actuarial application.

With regard to the two robust estimators $MTM$ and $MWM$, the 
placement of $\mathcal{O}$ at $d = 15$ causes fluctuation for 
all estimates of  $\mu, \text{VaR}_{0.99}, \text{TVaR}_{0.99}$ 
and $\text{PHDRM}_{0.99}$. That is because the chosen $a$ almost 
``break down'' against the left truncation point $d$. On the 
other hand, since all of the MTM and MWM estimators are designed 
to be sufficiently robust against upper data, we observe the 
stability of each corresponding risk measure from the right-hand 
side. Moreover, as trimming/winsorizing proportion $b$ increases, 
the stability range becomes wider. For example, $b=0.1$ maintain 
stability for about 57\% of the upper data range 
$(57\% \approx (1500 - 650)/(1500 - 15))$ while $b=0.15$ and 
$b=0.20$ have those  about 70\% and 84\%,  respectively. 
In particular, the premium pricing and risk control are 
unaffected when a new extreme claim (outlier) merges into 
the portfolio.
More importantly, with appropriate proportions $a$ and $b$ 
(i.e., $a=0.25, b=0.20$), $T3$ and $W3$ based values are 
closer moving toward and even coinciding with the theoretical 
targets (marked as dashed lines in Figure 
\ref{fig:PPZ_Graphs_RNG_35_PiP_99}).

From Figures 
\ref{fig:PPY_Graphs_RNG_83_PiP_99} and 
\ref{fig:PPZ_Graphs_RNG_35_PiP_99}, 
we clearly observed that MLE estimated 
statistics are highly volatile even for
truncated and/or censored dataset with 
sudden jumps at the censored thresholds.
A simple remedy for this volatile issue 
can be achieved via the robust MTM- or MWM-estimators.
These methods simply remove the impact of point masses 
accumulated by MLE at the censored thresholds.
For example, as seen in Table \ref{table:PPY_MLE_TW_ARE}
(see the entries inside the rectangular box),
out of $LN(1,4,2)$ distribution, 
if 10\% of the sample data are right censored 
and if we simply winsorize those 10\% censored data
values, i.e., $a=0,\ b = 10\%$ then the sensitivity 
curves turn out to be more stable and at the same 
time the asymptotic relative efficiency is
still maintained at 99.90\% level. 
The corresponding asymptotic relative efficiency 
for MTM is maintained at 94.20\% which is still 
close to 1.

All above - mentioned trends of sensitivity behavior 
(Figures \ref{fig:PPY_Graphs_RNG_83_PiP_99} and 
\ref{fig:PPZ_Graphs_RNG_35_PiP_99}) 
do not depend on the risk level.
That means, if we switch the risk threshold from
$p = 0.99$ to $p = 0.95$
or some other level, the corresponding
risk measure statistics will change but the 
fundamental pattern of sensitivity 
curves still remain the same.

\section{Real Data Illustrations}
\label{sec:RealDataAnalysis}

In this section, we apply the MWM, MTM and MLE
(both stand-alone lognormal as well as 
two composite models)
to analyze 1500 
indemnity losses in the United States provided by Insurance 
Service Office, Inc, which has been widely studied in 
the actuarial literature 
\citep[see, e.g.,][]{MR1988432,MR2108013,MR4118953} and are 
available in the R package 
\verb|copula| with data name \verb|loss|,
\cite{ky10}.
Our goal is to investigate what effect initial assumptions and
parameter estimation methods have on model fit and corresponding
insurance contract pricing.

\begin{figure}[hbt!]
\centering
 \includegraphics[width=1.05\textwidth]
{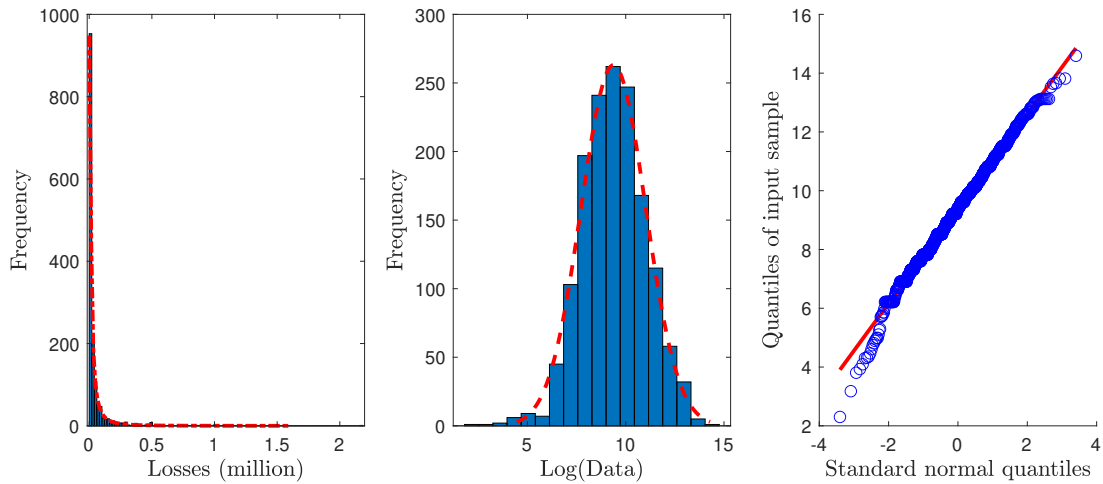}
\caption{The histograms of the top 1500 indemnity losses.
Left panel: Original data. 
Middle panel: Log-transformed data.  
Right panel: Log-normal QQ plot.}
\label{fig:real}
\end{figure}

A preliminary diagnostics (see in Figure \ref{fig:real}), which 
we have based on the histogram and normal QQ-plot of the 
log-transformed losses, shows that lognormal distribution may 
provide an adequate fit to indemnity losses. 
Thus, the usage of developed composite as well as 
adaptive models are illustrated on this loss 
data and the impact of trimming/winsorizing proportions on 
the lognormal model performance will be examined thoroughly.

\begin{table}[htb]
\centering
\caption{Payment-per-payment $Y$, payment-per-loss $Z$, 
transformation, and number of loss amounts with data 
truncation and censoring for the U.S. indemnity losses.}
\begin{tabular}{c|cc|cc|ccccc}
\hline  
\multirow{2}{*}{ Payment Type}  & 
\multicolumn{2}{c|}{Left Truncation} &  
\multicolumn{2}{c|}{Right Censoring} & 
\multicolumn{4}{c}{\# of loss amounts within each range}   \\
\cline{2-9}
{} & $d$ & $t$& $u$&$T$&$(0, d]$ & $(d, u)$ & $[u,\infty)$ & Total \\
\hline
$Y$  & 500  &6.2146 &  $10^5$ & 11.5129 & - & 1299 & 152 & 1451\\
\hline
$Z$ & 500  &6.2146 &  $10^5$ & 11.5129 & 49 & 1299 & 152 & 1500\\
\hline
\end{tabular}
\label{tab:Real_data} \\[5pt]
{\sc Note:} $t = \log (d - w_{0}),$ and $T = \log (u- w_{0})$.
\end{table}

We consider estimation of the loss severity component of 
the pure premium for an insurance benefit that equals to 
the amount by which an indemnity loss ($W$) exceeds 500 
(deductible, $d = 500$) dollars with a maximum benefit 
of 100,000  dollars (policy limit, $u=10^5$). 
Without loss of generality, we assume that $c=1$ since 
the asymptotic variances of all the estimators investigated 
in this paper do not depend on the coinsurance factor $c$. 
The corresponding lognormal payment-per-payment and 
payment-per-loss variables have the formulas as defined 
by $F_{Y}$ and $f_{Y}$, $F_{Z}$ and $f_{z}$, respectively, 
from \cite{MR4263275}, Section 2. Their transformations, 
and number of loss observations within each range of 
contract are also summarized in Table \ref{tab:Real_data}. 

Without considering the robustness, 
we first compare the performance of
three candidate models -- stand-alone lognormal,
composite \verb|LNPaI|, and composite \verb|LNGPD|
(mentioned in Section \ref{sec:CompositeModel}),
and determine the benchmark for further study of 
adaptive robust estimators. In this stage, 
all the parameters are estimated via
maximum likelihood estimation
and the corresponding statistics and performance 
indicators (such as Negative log-likelihood and
Akakie Information Criterion) are illustrated in 
Table \ref{table:CM_PPYZ}.

\begin{table}[hbt!]
\centering
{\small 
\caption{
MLE fitted composite models with some selection criterion
for the 1500 US indemnity payment-per-payment, $Y$, 
and payment-per-loss, $Z$, data scenarios.}
\label{table:CM_PPYZ}
\begin{tabular}{c|c|ccccc|c|cc|c}
\hline 
& & & & & & & & & & \\[-2.25ex] 
\multirow{1}{*}{\small Payments} & 
\multirow{1}{*}{\small Estimators} & 
\multirow{1}{*}{\small $\widehat{\theta}$} & 
\multirow{1}{*}{\small $\widehat{\sigma}$} &
\multirow{1}{*}{\small $\widehat{x}_{0}$} & 
\multirow{1}{*}{\small $\widehat{\alpha}$} &
\multirow{1}{*}{\small $\widehat{\lambda}$} &
\multirow{1}{*}{\small $w$} &
\multirow{1}{*}{\small NLL} &
\multirow{1}{*}{\small AIC} &
\multirow{1}{*}{\small LEV} \\
\hline\hline
\multirow{3}{*}{$Y$} & 
\verb|LN| & 
9.43 & 1.59 & - & - & - & 1 & 14,456.28 & 28,916.55 & {2.675} \\
\cline{2-11}
{} & \verb|LNPaI| & 
9.35 & 1.49  & 82,282 &0.885 & - & 0.878 &  14,454.19 & 28,914.37 & {2.661} \\
\cline{2-11}
{} & \verb|LNGDP| & 
9.33 & 1.47& 14,244 & 1.020 & 11,700  &0.541 & 14,454.22 & 28,916.44 & {2.663} \\
\hline\hline 
\multirow{3}{*}{$Z$} & 
\verb|LN| & 
9.39 & 1.64 & - & - & - & 1 & 14,674.03 & 29,352.06 & 2.600 \\
\cline{2-11}
{} & \verb|LNPaI| & 
9.30 & 1.58  & 99,999 &0.883 & - & 0.895 & 14,673.95 & 29,353.90 & {2.595} \\
\cline{2-11}
{} & \verb|LNGDP| & 
9.92  & 1.83& 4,239 & 1.027 & 11,961  &0.258 & 14,673.44 & 29,354.87 & 2.582 \\
\hline
\end{tabular} \\[5pt]
}
{\sc Note:} 
$x_{0}$: 
Location that separates Lognormal and Pareto models.
$w$: proportion of Lognormal distribution.
NLL: Negative log-likelihood. AIC: AIC criterion.
The limited expected value LEV is in $10^4$.
\end{table}

Obviously,
there is not much improvement either in NLL 
or in AIC going from stand-alone lognormal 
model to either \verb|LNPaI| or to \verb|LNGPD|. 
The percentage changes are clearly way smaller 
than $0.50\%$. 
Therefore, at least out of this data illustration,
what we can conclude is that for payment-per-payment
and payment-per-loss, that is, for truncated and/or
censored loss data scenarios, the stand-alone
lognormal model is equally powerful to model 
those data scenarios as the competing composite 
models. 
This is because the composite models are specially
more powerful to provide a good fit to the entire
range of the loss data given that there are large
losses with small probabilities, but for payments
$Y$ and $Z$ those large losses are already
censored at the policy limit, $u$. 
Hence, due to the simplicity of the distribution,
the stand-alone lognormal model is the optimal 
option to serve as the MLE benchmark in the
following step of methodology comparisons. 

Next, considering model robustness, 
we discuss the model fit of the MTM 
and MWM estimators in $PPY$ and $PPZ$ 
scenarios when policy deductible and 
policy limit are included in the risk pricing.
In Section \ref{sec:Introduction},
we mentioned that MWM for truncated and censored
loss data is an adaptive estimation procedures. 
This is the right place to describe why we 
call MWM an adaptive estimation procedure. 
First, for payment $Y$, the MWM estimators as well
as their asymptotic results are based on the assumption
of $0 \le a < 1-b \le s^{*} \le 1$. 
Then, an obvious challenge 
in estimation procedures is to maintain the 
condition $0 \le a < 1-b \le s^{*} \le 1$ 
both empirically and parametrically.
Initially, we set
\[
\hat{s}_{\mbox{\tiny E}}^{*}
:=
\dfrac{F_{n}(T)-F_{n}(t)}{1-F_{n}(t)}
=
{n_{1}}/{n},
\quad 
\mbox{where $F_{n}$ is the empirical cdf},
\]
and choose $b$ such that 
$1-b \le \hat{s}_{\mbox{\tiny E}}^{*}$.
But after estimating 
$\widehat{\theta}_{\mbox{\tiny y,MWM}}$ and 
$\widehat{\sigma}_{\mbox{\tiny y,MWM}}$,
it could be the case that 
\[
\hat{s}_{\mbox{\tiny P}}^{*}
:=
\dfrac{\widehat{F}_{X}(T)-\widehat{F}_{X}(t)}
{1-\widehat{F}_{X}(t)}
< 1-b,
\quad 
\mbox{where $\widehat{F}_{X}$ is the estimated parametric cdf},
\]
which is unsatisfactory for statistical inferences as 
it violates the assumption 
(Case II from Section \ref{sec:PPY_MWM}).
Therefore, the value of $b$ should be chosen 
{\em adaptively}
such that  
$1-b \le 
\min\{
\hat{s}_{\mbox{\tiny E}}^{*},
\hat{s}_{\mbox{\tiny P}}^{*}
\}.
$
Likewise, for payment $Z$, the left and right trimming proportions
$a$ and $b$ are also chosen adaptively:
\[
\max \{F_{n}(t),\widehat{F}_{X}(t)\}
\le 
a
\quad \mbox{and} \quad 
1-b 
\le
\min \{F_{n}(T),\widehat{F}_{X}(T)\}.
\]
The trimming and winsorizing proportions in Tables \ref{table:estTablePymtY}
and \ref{table:estTablePymtZ} are chosen accordingly.
Adaptive MTM and MWM are now, respectively, called 
by AMTM and AMWM for the rest of this section.

As observed in the indemnity data set, 49 losses among 1500 claims 
are below the deductible, and 152 losses are above the policy limit. 
Since our major objective is to see if the fitted model captures 
the behavior of the underlying distribution, it is natural to choose 
$a$ and $b$ proportions that fluctuate across left-truncation 
(deductible) and right censored (policy limit), separately. 
In addition, to see the impact of chosen $b$ at the policy 
threshold ($u$) alone, we  fix $a=0$ and set different values of 
$b$ from small $150/1451$ to large $750/1451$.  The estimates of 
location parameter $\theta$ and scale parameter $\sigma$ for 
lognormal model are provided and the corresponding 95\% 
confidence intervals are built. Further, in Tables  
\ref{table:estTablePymtY} and \ref{table:estTablePymtZ}, 
we also provide the actuarial premiums (or call it limited 
expected value (LEV)) calculated as 
\begin{itemize}
\item 
$\text{LEV}=\dfrac{E[W\wedge u]-E[W\wedge d]}{1-F_{W}(d)}$ 
for payment-per-payment; and
\item $\text{LEV}=E[W\wedge u]-E[W\wedge d]$ 
for payment-per-loss,
\end{itemize}
where
$$E[W \wedge w] = w_{0} 
+ e^{\theta + \frac{\sigma^2}{2}}
\Phi\left(\dfrac{\ln{(w-w_{0})}-\theta-\sigma^2}{\sigma} \right) 
+ (w-w_{0})\left[1-\Phi \left( 
\dfrac{\ln{(w-w_{0})}-\theta}{\sigma}\right)\right].
 $$
Moreover, Asymptotic Relative Efficiency (ARE) and 
Kolmogorov–Smirnov test (KS Test) on each base are
utilized to compare the model robustness and fit
accuracy of various AMWM, AMTM and MLE estimators.

\begin{table}[hbt!]
\centering
{\small
\caption{
MLE, AMTM and AMWM estimators of $\theta$ and $\sigma$ with 
their corresponding asymptotic confidence intervals and 
estimated values of AREs for payment $Y$ data scenario.
}
\label{table:estTablePymtY}
\begin{tabular}{c|cc|cc|cc|cc|c|c|cc}
\cline{2-13}
{}&
\multicolumn{2}{|c|}{\multirow{2}{*}{Estimators}}
 & 
\multirow{2}{*}{$\hat{\theta}$} & 
\multirow{2}{*}{$\hat{\sigma}$} &
\multicolumn{2}{c|}{$\hat{s}^{\ast}$} &
\multicolumn{2}{c|}{95\% CI for}&
\multirow{2}{*}{LEV}&
\multirow{2}{*}{$\widehat{ARE}$}& \multicolumn{2}{c}{KS Test}\\
\cline{6-9} \cline{12-13}
& & & & & $\hat{s}_{E}^{\ast}$ & $\hat{s}_{P}^{\ast}$& $\theta$& 
$\sigma$& & & $D_{y}$ & $h_{y}$  \\
\hline\hline
\multicolumn{3}{c|}{MLE} & 
9.43 & 1.59 & - & - & (9.34, 9.52)
 & (1.52, 1.67)& 2.675& 1.00 & 0.032 & 0 \\
 \hline\hline
 {}& $a$ & $b$ &\multicolumn{9}{c}{Condition required: 
 $1-b\leq \min\{\hat{s}_{E}^{\ast}, \hat{s}_{P}^{\ast}\}$}\\
 \hline
 \multirow{8}{*}
{\rotatebox{90}{AMWM}} 
& 0.00& 150/1451 & 9.43 & 1.59 &0.90&0.90 & (9.34, 9.52)
 & (1.51, 1.67)& {2.671}& 0.99 & 0.033& 0\\
 
{} & 0.00& 200/1451 & 9.43 & 1.58 &0.90&0.90 & (9.34, 9.52)
 & (1.50, 1.66)& {2.664}& 0.95 & 0.033& 0\\
 
 {} & 0.00& 300/1451 & 9.43 & 1.57 &0.90&0.91 & (9.34, 9.52)
 & (1.49, 1.66)& {2.656}& 0.88 & 0.034& 0\\
 
 {} & 0.00& 700/1451 & 9.45 & 1.58 &0.90&0.90 & (9.35, 9.55)
 & (1.46, 1.71)& {2.701}& 0.57 & 0.038& 1\\
\cline{2-13}
{} & 10/1451& 150/1451 & 9.43 & 1.59 &0.90&0.91 & (9.34, 9.52)
 & (1.51, 1.66)& {2.671}& 0.99 & 0.033& 0\\
 
{} & 50/1451& 200/1451 & 9.42 & 1.60 &0.90&0.91 & (9.33, 9.51)
 & (1.52, 1.69)& {2.672}& 0.95 & 0.030& 0\\
 
{} & 100/1451& 300/1451 & 9.42 & 1.60 &0.90&0.90 & (9.32, 9.51)
 & (1.51, 1.69)& {2.670}& 0.86 & 0.029& 0\\
 
{} & 650/1451& 650/1451 & 9.37 & 1.61 &0.90&0.91 & (9.25, 9.48)
 & (1.35, 1.91)& {2.598}& 0.24 & 0.031& 0\\
 \hline\hline

 \multirow{8}{*}
{\rotatebox{90}{AMTM}} 
& 0.00& 150/1451 & 9.42 & 1.56 &0.90&0.91 & (9.34, 9.51)
 & (1.49, 1.65)& {2.634}& 0.94 & 0.034& 0\\
 
{} & 0.00& 200/1451 & 9.42 & 1.55 &0.90&0.91 & (9.33, 9.51)
 & (1.47, 1.64)& {2.618}& 0.89 & 0.034& 0\\
 
 {} & 0.00& 300/1451 & 9.42 & 1.54 &0.90&0.91 & (9.33, 9.50)
 & (1.45, 1.63)& {2.591}& 0.80 & 0.034& 0\\
 
 {} & 0.00& 700/1451 & 9.37 & 1.47 &0.90&0.93 & (9.27, 9.47)
 & (1.35, 1.59)&{2.418}&  0.48 & 0.043& 1\\
\cline{2-13}
{} & 10/1451& 150/1451 & 9.42 & 1.57 &0.90&0.91 & (9.33, 9.51)
 & (1.49, 1.65)& {2.637}& 0.94 & 0.033& 0\\
 
{} & 50/1451& 200/1451 & 9.41 & 1.59 &0.90&0.91 & (9.32, 9.50)
 & (1.50, 1.67)&{2.640}&  0.89 & 0.030& 0\\
 
{} & 100/1451& 300/1451 & 9.40 & 1.59 &0.90&0.90 & (9.31, 9.50)
 & (1.50, 1.69)& {2.639}& 0.79 & 0.028& 0\\
 
{} & 650/1451& 650/1451 & 9.26 & 2.09 &0.90&0.85 & (8.96, 9.56)
 & (1.56, 2.81)& {3.038}& 0.24 & 0.064& 1\\
\hline
\end{tabular} \\[5pt]
{
\small 
{\sc Note}
: The limited expected value LEV is in $10^4$.
}
}
\end{table}

As Tables \ref{table:estTablePymtY} and \ref{table:estTablePymtZ} 
suggest, the AMWM and AMTM estimators with appropriate proportions
$a$ and $b$ (i.e. $a=0, b=150/1451$ in Table \ref{table:estTablePymtY} 
and $a=75/1500, b=150/1500$ in Table \ref{table:estTablePymtZ}) lead 
to premium estimates that are close enough to that of non-robust 
estimator MLE. 
On the other hand, the scale estimate, confidence 
interval and LEV of AMWM and AMTM estimators that obtained with highly 
robust but inefficient estimators (i.e., $b=700/1451$ or $b=650/1451$) 
have a significant deviation from the corresponding MLE. 
In terms of ARE, 
the values of payment $Y$ depend on the unknown parameters of $\theta$ 
and $\sigma$ while those of payment $Z$ do not, resulting in the degree 
of efficiency loss varies from the former to the latter.

In addition, we note the remarkable stability of point estimates, 
interval estimates as well as LEV premiums based on small proportions 
of adaptive trimming/winsorizing when the assumed distribution is lognormal. 
Besides, we again observe substantial differences between 
AMWM and
AMTM fits, in particular, for large $a$ and $b$ proportions. 
For example, in Table \ref{table:estTablePymtY},  
when $a = b = 650/1451$, AMWM has LEV$ = 2.598 \times 10^4$ 
whereas AMTM has LEV$ = 3.038 \times 10^4$. 
However, the ARE 
of these two estimators are almost identical, which shows that 
even in the worst case, the AMWM is still better than AMTM for 
pricing risks.

\begin{table}[hbt!]
\centering
{\small 
\caption{
MLE, AMTM and AMWM estimators of $\theta$ and $\sigma$ with 
their corresponding asymptotic confidence intervals and 
estimated values of AREs for payment $Z$ data scenario.
}
\label{table:estTablePymtZ}
\begin{tabular}{c|cc|cc|cc|cc|c|c|cc}
\cline{2-13}
{}&
\multicolumn{2}{|c|}{\multirow{2}{*}{Estimators}}
 & 
\multirow{2}{*}{$\hat{\theta}$} & 
\multirow{2}{*}{$\hat{\sigma}$} &
\multirow{2}{*}{$\widehat{F}_{X}(t)$} &
\multirow{2}{*}{$\widehat{F}_{X}(T)$} & 
\multicolumn{2}{c|}{95\% CI for}&
\multirow{2}{*}{LEV}&
\multirow{2}{*}{$\widehat{ARE}$}& \multicolumn{2}{c}{KS Test}\\
\cline{8-9} \cline{12-13}
& & & & &  & & $\theta$& $\sigma$&  & & $D_{z}$ & $h_{z}$ \\
\hline\hline
\multicolumn{3}{c|}{MLE} & 
9.39 & 1.64 & - & - & (9.30, 9.47)
& (1.58, 1.71)& {2.600}& 1.00 & 0.027 & 0 \\
\hline\hline
{}& \multirow{2}{*}{$a$} & \multirow{2}{*}{$b$} &
\multicolumn{9}{c}{Condition required:}\\
{}&  &  &\multicolumn{9}{c}{ $\max\{\widehat{F}_{n}(t),
\widehat{F}_{X}(T)\}\leq a$ and $1-b\leq \min\{\widehat{F}_{n}(t),\widehat{F}_{X}(T)\}$.}\\
\hline
\multirow{8}{*}
{\rotatebox{90}{AMWM}} 
& 75/1500& 150/1500 & 9.40 & 1.61 &0.02&0.91 & (9.32, 9.48)
& (1.54, 1.67)& {2.585}&0.97 & 0.031& 0\\

{} & 75/1500& 225/1500 & 9.39 & 1.60 &0.02&0.91 & (9.31, 9.48)
& (1.53, 1.67)& {2.567}&0.93 & 0.031& 0\\

{} & 75/1500& 375/1500 & 9.38 & 1.58 &0.02&0.91 & (9.30, 9.47)
& (1.51, 1.66)& {2.533}&0.83 & 0.031& 0\\

{} & 75/1500& 750/1500 & 9.38 & 1.57 &0.02&0.91 & (9.28, 9.48)
& (1.48, 1.67)& {2.519}&0.59 & 0.031& 0\\
\cline{2-13}
{} & 150/1500& 150/1500 & 9.39 & 1.63 &0.03&0.90 & (9.30, 9.47)
& (1.56, 1.70)& {2.592}&0.93 & 0.026& 0\\

 {} &225/1500& 225/1500 & 9.39 & 1.62 &0.03&0.90 & (9.30, 9.47)
 & (1.55, 1.70)& {2.578}&0.83 & 0.027& 0\\
 
  {} &375/1500& 375/1500 & 9.38 & 1.61 &0.02&0.91 & (9.29, 9.47)
 & (1.52, 1.70)& {2.552}&0.64 & 0.027& 0\\
 
  {} &700/1500& 700/1500 & 9.40 & 2.26 &0.08&0.82 & (9.26, 9.54)
 & (1.87, 2.74)& {3.140}&0.17 & 0.095& 1\\

\hline\hline

\multirow{8}{*}
{\rotatebox{90}{AMTM}} 
& 75/1500& 150/1500 & 9.38 & 1.62 &0.03&0.91 & (9.30, 9.47)
 & (1.55, 1.69)& {2.570}& 0.92 & 0.027& 0\\
 
{} & 75/1500& 225/1500 & 9.38 & 1.61 &0.02&  0.91 & (9.30, 9.47)
 & (1.54, 1.69)& {2.558}&0.86 & 0.027& 0\\
 
 {} & 75/1500& 375/1500 & 9.38 & 1.60 &0.02&0.91 & (9.29, 9.46)
 & (1.53, 1.69)&{2.544}& 0.76 & 0.027& 0\\
 
 {} & 75/1500& 750/1500 & 9.36 & 1.59 &0.02&0.91 & (9.26, 9.47)
 & (1.49, 1.70)&{2.506}& 0.52 & 0.028& 0\\
\cline{2-13}
 {} & 150/1500& 150/1500 & 9.38 & 1.63 &0.03&0.90 & (9.30, 9.47)
 & (1.55, 1.70)&{2.575}& 0.86 & 0.026& 0\\
 
 {} &225/1500& 225/1500 & 9.38 & 1.63 &0.03&0.90 & (9.29, 9.46)
 & (1.55, 1.72)& {2.573}&0.76 & 0.026& 0\\
 
  {} &375/1500& 375/1500 & 9.38 & 1.61 &0.02&0.91 & (9.29, 9.47)
 & (1.50, 1.71)&{2.551}& 0.57 & 0.027& 0\\
 
  {} &700/1500& 700/1500 & 9.38 & 2.36 &0.09&0.82 & (9.23, 9.52)
 & (1.92, 2.91)& {3.172}&0.16 & 0.107& 1\\
 
\hline
\end{tabular} \\[5pt]
{
\small 
{\sc Note:}
The limited expected value LEV is in $10^4$.
}
}
\end{table}

For hypothesis testing purpose for the various fitted models,
we use the Kolmogorov-Smirnov (KS) test statistic. 
The KS test statistic for 
payments $Y$ and $Z$ data sets are, respectively,
defined as \citep[see, e.g.,][p. 360]{MR3890025}:
\begin{align}
D_{y}
& :=
\max_{0 < y
\leq 
c(T-t)}
\left|F_{n}(y)-\widehat{F}_{Y}(y)\right|; 
\label{eqn:KS_PPY_Defn} \\
\label{eqn:KS_PPZ_Defn}
D_{z}
& :=
\max_{0 \le z
\leq 
c(T-t)
}
\left|F_{n}(z)-\widehat{F}_{Z}(z)\right|,
\end{align}
where $\widehat{F}_{Y}$ and $\widehat{F}_{Z}$
are the the estimated cdf of 
$F_{Y}$ and $F_{Z}$, respectively.
For each fitted model, the corresponding
KS test statistics $\{D_{y},D_{z}\}$ 
and the decision $\{h_{y},h_{z}\}$
of the hypothesis test are given in the 
last two columns of Tables 
\ref{table:estTablePymtY} and 
\ref{table:estTablePymtZ}
where $h_{y},h_{z} = 0$ indicates that the lognormal model
is plausible and $h_{y},h_{z} = 1$ means that the 
the lognormal model is rejected at the 
significance level of $5\%$.

\section{Conclusion}
\label{sec:Conclusion}

In this paper, we have introduced and developed a new estimation
procedure -- method of winsorized moments (MWM) -- for 
robust fitting of truncated and censored lognormal distributions,
which are appropriate for insurance payment-per-payment and 
payment-per-loss data. 
Large-sample properties of the MWM estimators have 
been established, and their small-sample performance has been 
investigated through simulations and compared to that of the main 
competitor (method of trimmed moments, MTM) and the 
benchmark MLE.
Among the three methods, MLE is not robust whereas 
MTM and MWM are both robust and exhibit similar robustness 
properties. 
In terms of efficiency, MLE is most efficient, of course, 
and MWM consistently outperforms MTM under the lognormal model 
with policy deductible and limit. 
Moreover, as the sample size 
increases, finite-sample measures of the estimators' performance 
approach their asymptotic counterparts.
Further, a sensitivity study has been conducted to gauge MWM's, 
MTM's, and MLE's response to a single outlier (new extreme 
observation) that is placed at various locations within the 
sample ranging from the point of left-truncation (deductible) to 
that of right-censoring (policy limit). It has been found that 
MLE-based sensitivity curves are influenced by all locations of 
$\mathcal{O}$ with the most dramatic effect when outlier is 
placed at the censoring threshold. 
Finally, the numerical examples based on 1500 U.S. indemnity losses 
have revealed that evaluation of the net premiums using adaptively 
computed right-trimming and right-winsorizing proportion for 
AMTM and AMWM, respectively, leads to reasonable results.
By using theses 1500 U.S. indemnity losses, 
it is also demonstrated that,
specifically for truncated and censored data,
the composite models do not provide much 
predictive power compared to the corresponding 
competing stand-alone severity model. 

\baselineskip 4.80mm
\bibliography{ArXiv}

\begin{thebibliography}{}

\bibitem[\protect\astroncite{Beirlant et~al.}{2004}]{MR2108013}
Beirlant, J., Goegebeur, Y., Teugels, J., and Segers, J. (2004).
\newblock {\em Statistics of Extremes: Theory and Applications}.
\newblock Wiley Series in Probability and Statistics. John Wiley \& Sons, Ltd.,
  Chichester.

\bibitem[\protect\astroncite{Blostein and Miljkovic}{2019}]{MR3896968}
Blostein, M. and Miljkovic, T. (2019).
\newblock On modeling left-truncated loss data using mixtures of distributions.
\newblock {\em Insurance: Mathematics \& Economics}, 85:35--46.

\bibitem[\protect\astroncite{Brazauskas et~al.}{2009}]{MR2497558}
Brazauskas, V., Jones, B.~L., and Zitikis, R. (2009).
\newblock Robust fitting of claim severity distributions and the method of
  trimmed moments.
\newblock {\em Journal of Statistical Planning and Inference},
  139(6):2028--2043.

\bibitem[\protect\astroncite{Brazauskas and Kleefeld}{2016}]{MR3474025}
Brazauskas, V. and Kleefeld, A. (2016).
\newblock Modeling severity and measuring tail risk of {N}orwegian fire claims.
\newblock {\em North American Actuarial Journal}, 20(1):1--16.

\bibitem[\protect\astroncite{Chernoff et~al.}{1967}]{MR0203874}
Chernoff, H., Gastwirth, J.~L., and Johns, Jr., M.~V. (1967).
\newblock Asymptotic distribution of linear combinations of functions of order
  statistics with applications to estimation.
\newblock {\em Annals of Mathematical Statistics}, 38(1):52--72.

\bibitem[\protect\astroncite{Cohen}{1950}]{MR0038041}
Cohen, Jr., A.~C. (1950).
\newblock Estimating the mean and variance of normal populations from singly
  truncated and doubly truncated samples.
\newblock {\em Annals of Mathematical Statistics}, 21(4):557--569.

\bibitem[\protect\astroncite{Cohen}{1951}]{MR0045361}
Cohen, Jr., A.~C. (1951).
\newblock On estimating the mean and variance of singly truncated normal
  frequency distributions from the first three sample moments.
\newblock {\em Annals of the Institute of Statistical Mathematics}, 3:37--44.

\bibitem[\protect\astroncite{Cooray and Ananda}{2005}]{MR2188658}
Cooray, K. and Ananda, M. M.~A. (2005).
\newblock Modeling actuarial data with a composite lognormal-{P}areto model.
\newblock {\em Scandinavian Actuarial Journal}, 2005(5):321--334.

\bibitem[\protect\astroncite{Delong et~al.}{2021}]{MR4340261}
Delong, L.~u., Lindholm, M., and W\"{u}thrich, M.~V. (2021).
\newblock Gamma mixture density networks and their application to modelling
  insurance claim amounts.
\newblock {\em Insurance: Mathematics \& Economics}, 101:240--261.

\bibitem[\protect\astroncite{Fabi\'{a}n}{2001}]{MR1862941}
Fabi\'{a}n, Z. (2001).
\newblock Induced cores and their use in robust parametric estimation.
\newblock {\em Communications in Statistics: Theory and Methods},
  30(3):537--555.

\bibitem[\protect\astroncite{Fabi\'{a}n}{2008}]{MR2412617}
Fabi\'{a}n, Z. (2008).
\newblock New measures of central tendency and variability of continuous
  distributions.
\newblock {\em Communications in Statistics: Theory and Methods},
  37(1-2):159--174.

\bibitem[\protect\astroncite{Fabi\'{a}n}{2010}]{Fab10}
Fabi\'{a}n, Z. (2010).
\newblock Scalar score function and score correlation.
\newblock Technical Report 1077, Institute of Computer Science, Academy of
  Sciences of the Czech Republic.

\bibitem[\protect\astroncite{Frees}{2017}]{MR3765327}
Frees, E. (2017).
\newblock Insurance portfolio risk retention.
\newblock {\em North American Actuarial Journal}, 21(4):526--551.

\bibitem[\protect\astroncite{Frees and Valdez}{1998}]{MR1988432}
Frees, E.~W. and Valdez, E.~A. (1998).
\newblock Understanding relationships using copulas.
\newblock {\em North American Actuarial Journal}, 2(1):1--25.

\bibitem[\protect\astroncite{Fung}{2021}]{tcf21}
Fung, T.~C. (2021).
\newblock Maximum weighted likelihood estimator for robust heavy-tail modelling
  of finite mixture models.
\newblock {\em Available at ArXiv.org/abs/2108.01356}.

\bibitem[\protect\astroncite{Goffard and Laub}{2021}]{MR4287445}
Goffard, P.-O. and Laub, P.~J. (2021).
\newblock Approximate {B}ayesian computations to fit and compare insurance loss
  models.
\newblock {\em Insurance: Mathematics \& Economics}, 100:350--371.

\bibitem[\protect\astroncite{Gui et~al.}{2021}]{MR4163087}
Gui, W., Huang, R., and Lin, X.~S. (2021).
\newblock Fitting multivariate {E}rlang mixtures to data: a roughness penalty
  approach.
\newblock {\em Journal of Computational and Applied Mathematics}, 386:113216,
  17.

\bibitem[\protect\astroncite{Hewitt et~al.}{1979}]{hl79}
Hewitt, C.~C., Jr., and Lefkowitz, B. (1979).
\newblock Methods for fitting distributions to insurance loss data.
\newblock In {\em Proceedings of the Casualty Actuarial Society}, volume LXVI,
  pages 139--160. Casualty Actuarial Society, VA.

\bibitem[\protect\astroncite{Klugman et~al.}{2019}]{MR3890025}
Klugman, S.~A., Panjer, H.~H., and Willmot, G.~E. (2019).
\newblock {\em Loss Models: From Data to Decisions}.
\newblock John Wiley \& Sons, Hoboken, NJ, fifth edition.

\bibitem[\protect\astroncite{Kojadinovic and Yan}{2010}]{ky10}
Kojadinovic, I. and Yan, J. (2010).
\newblock Modeling multivariate distributions with continuous margins using the
  copula r package.
\newblock {\em Journal of Statistical Software}, 34(9):1--20.

\bibitem[\protect\astroncite{Michael et~al.}{2020}]{MR4118953}
Michael, S., Miljkovic, T., and Melnykov, V. (2020).
\newblock Mixture modeling of data with multiple partial right-censoring
  levels.
\newblock {\em Advances in Data Analysis and Classification. ADAC},
  14(2):355--378.

\bibitem[\protect\astroncite{Miljkovic and Gr\"{u}n}{2016}]{MR3543061}
Miljkovic, T. and Gr\"{u}n, B. (2016).
\newblock Modeling loss data using mixtures of distributions.
\newblock {\em Insurance: Mathematics \& Economics}, 70:387--396.

\bibitem[\protect\astroncite{Nadarajah and Bakar}{2014}]{MR3177097}
Nadarajah, S. and Bakar, S. A.~A. (2014).
\newblock New composite models for the {D}anish fire insurance data.
\newblock {\em Scandinavian Actuarial Journal}, 2014(2):180--187.

\bibitem[\protect\astroncite{Pigeon and Denuit}{2011}]{MR2842558}
Pigeon, M. and Denuit, M. (2011).
\newblock Composite lognormal-{P}areto model with random threshold.
\newblock {\em Scandinavian Actuarial Journal}, 2011(3):177--192.

\bibitem[\protect\astroncite{Poudyal}{2021a}]{MR4263275}
Poudyal, C. (2021a).
\newblock Robust estimation of loss models for lognormal insurance payment
  severity data.
\newblock {\em ASTIN Bulletin}, 51(2):475--507.

\bibitem[\protect\astroncite{Poudyal}{2021b}]{MR4192140}
Poudyal, C. (2021b).
\newblock Truncated, censored, and actuarial payment-type moments for robust
  fitting of a single-parameter {P}areto distribution.
\newblock {\em Journal of Computational and Applied Mathematics}, 388:113310,
  18.

\bibitem[\protect\astroncite{Punzo et~al.}{2018}]{MR3836650}
Punzo, A., Bagnato, L., and Maruotti, A. (2018).
\newblock Compound unimodal distributions for insurance losses.
\newblock {\em Insurance: Mathematics \& Economics}, 81:95--107.

\bibitem[\protect\astroncite{Scollnik}{2007}]{MR2347211}
Scollnik, D. P.~M. (2007).
\newblock On composite lognormal-{P}areto models.
\newblock {\em Scandinavian Actuarial Journal}, 2007(1):20--33.

\bibitem[\protect\astroncite{Serfling}{2002}]{MR1987777}
Serfling, R. (2002).
\newblock Efficient and robust fitting of lognormal distributions.
\newblock {\em North American Actuarial Journal}, 6(4):95--109.

\bibitem[\protect\astroncite{Serfling}{1980}]{MR595165}
Serfling, R.~J. (1980).
\newblock {\em Approximation Theorems of Mathematical Statistics}.
\newblock John Wiley \& Sons, New York.

\bibitem[\protect\astroncite{Shah and Jaiswal}{1966}]{MR0196848}
Shah, S.~M. and Jaiswal, M.~C. (1966).
\newblock Estimation of parameters of doubly truncated normal distribution from
  first four sample moments.
\newblock {\em Annals of the Institute of Statistical Mathematics},
  18:107--111.

\bibitem[\protect\astroncite{Stehl\'{\i}k et~al.}{2010}]{MR2720398}
Stehl\'{\i}k, M., Potock\'{y}, R., Waldl, H., and Fabi\'{a}n, Z. (2010).
\newblock On the favorable estimation for fitting heavy tailed data.
\newblock {\em Computational Statistics}, 25(3):485--503.

\bibitem[\protect\astroncite{Tomarchio and Punzo}{2020}]{MR4149559}
Tomarchio, S.~D. and Punzo, A. (2020).
\newblock Dichotomous unimodal compound models: application to the distribution
  of insurance losses.
\newblock {\em Journal of Applied Statistics}, 47(13-15):2328--2353.

\bibitem[\protect\astroncite{Tukey}{1960}]{MR0120720}
Tukey, J.~W. (1960).
\newblock A survey of sampling from contaminated distributions.
\newblock In {\em Contributions to Probability and Statistics}, pages 448--485.
  Stanford University Press, Stanford, CA.

\bibitem[\protect\astroncite{van~der Vaart}{1998}]{MR1652247}
van~der Vaart, A.~W. (1998).
\newblock {\em Asymptotic Statistics}.
\newblock Cambridge University Press, Cambridge.

\bibitem[\protect\astroncite{Zhao et~al.}{2018a}]{MR3758788}
Zhao, Q., Brazauskas, V., and Ghorai, J. (2018a).
\newblock Robust and efficient fitting of severity models and the method of
  {W}insorized moments.
\newblock {\em ASTIN Bulletin}, 48(1):275--309.

\bibitem[\protect\astroncite{Zhao et~al.}{2018b}]{MR3750626}
Zhao, Q., Brazauskas, V., and Ghorai, J. (2018b).
\newblock Small-sample performance of the {MTM} and {MWM} estimators for the
  parameters of log-location-scale families.
\newblock {\em Journal of Statistical Computation and Simulation},
  88(4):808--824.

\end{thebibliography}
\thispagestyle{plain}
\pagestyle{plain} 
\thispagestyle{plain}

\begin{appendices}

\section{Auxiliary Results}
\label{sec:AppendixA}

Recall that, 
for $0 \le s \le 1$, we define
$\bar{s} = 1-s$ and
$\Delta_{s} := \Phi^{-1}\left(s+\bar{s}\Phi(\gamma)\right)$.

\begin{enumerate}[label=\Roman*.]
\item 
The expressions for $c_{y,k}$, $1 \le k \le 4$
used in Section \ref{sec:PPY_MWM} are given by:
\[
c_{y,k} 
\equiv
c_{y, k}(\Phi,a,b,\gamma) 
=
a
\Delta_{a}^{k}
+ \int_{a}^{\bar{b}}
\Delta_{s}^{k} \, ds 
+ b
\Delta_{\bar{b}}^{k},
\]
and consequently the corresponding expressions 
for $c_{k}$ used in Section \ref{sec:PPZ_MWM}
are computed as follows:
\[
c_{k} 
\equiv 
c_{k}(\Phi,a,b) 
=
\lim_{\gamma \to -\infty} 
c_{y,k} 
=
a
\left[ 
\Phi^{-1}
\left(
a
\right)
\right]^{k}
+ \int_{a}^{\bar{b}}
\left[ 
\Phi^{-1}
\left(
s
\right)
\right]^{k} \, ds 
+ b
\left[ 
\Phi^{-1}
\left(
\bar{b}
\right)
\right]^{k}.
\]

\item 
Following the appendix of \cite{MR3758788}, 
the expressions presented in the entries of the 
variance-covariance  matrix 
$\bf{\Sigma}_{y}$ from Section \ref{sec:PPY_MWM}
are as follows: 
\begin{align*}
c_{y,1}^{*}
& =
c_{y,2}-c_{y,1}^2
-a\dfrac{\partial (c_{y,2}-c_{y,1}^2)}{\partial a}
-b\dfrac{\partial (c_{y,2}-c_{y,1}^2)}{\partial b} 
+a\bar{a}
\left(
\dfrac{\partial c_{y,1}}{\partial a}
\right)^2
+b\bar{b}
\left(
\dfrac{\partial c_{y,1}}{\partial b}
\right)^2
-2ab\dfrac{\partial c_{y,1}}{\partial a}
\dfrac{\partial c_{y,1}}{\partial b}, \\[12pt]
2 c_{y,2}^{*}
& =
c_{y,3}-c_{y,1}\cdot c_{y,2}
-a\dfrac{\partial (c_{y,3}-c_{y,1}c_{y,2})}{\partial a}
-b\dfrac{\partial (c_{y,3}-c_{y,1}\cdot c_{y,2})}{\partial b} \\
& \quad 
+a\bar{a}\dfrac{\partial c_{y,1}}{\partial a}
\dfrac{\partial c_{y,2}}{\partial a}
+b\bar{b}\dfrac{\partial c_{y,1}}{\partial b}
\dfrac{\partial c_{y,2}}{\partial b}
-ab
\left(
\dfrac{\partial c_{y,1}}{\partial a}
\dfrac{\partial c_{y,2}}{\partial b}
+\dfrac{\partial c_{y,1}}{\partial b}\dfrac{\partial c_{y,2}}{\partial a}
\right), \\[12pt]
4 c_{y,3}^{*}
& =
c_{y,4}-c_{y,2}^2
-a\dfrac{\partial (c_{y,4}
-c_{y,2}^2)}{\partial a}
-b\dfrac{\partial (c_{y,4}-c_{y,2}^2)}{\partial b} 
+a\bar{a}
\left(
\dfrac{\partial c_{y,2}}{\partial a}
\right)^2
+b\bar{b}
\left(
\dfrac{\partial c_{y,2}}{\partial b}
\right)^2
-2ab\dfrac{\partial c_{y,2}}
{\partial a}\dfrac{\partial c_{y,2}}{\partial b},
\end{align*}
where the involved derivatives are listed below:
\begin{eqnarray*}
\dfrac{\partial c_{y,k}}{\partial a}
& = & 
\Delta_{a}^{k}
+\dfrac{ak\Delta_{a}^{k-1}}{\phi(\Delta_{a})}
\bar{\Phi}(\gamma)-\Delta_{a}^{k}
=
a\bar{\Phi}(\gamma)
\dfrac{k \Delta_{a}^{k-1}}{\phi(\Delta_{a})}, 
\qquad 
\dfrac{\partial c_{y,k}}{\partial b}
=
-b\bar{\Phi}(\gamma)
\dfrac{k\Delta_{\bar{b}}^{k-1}}
{\phi(\Delta_{\bar{b}})}, \\[5pt]
c_{y,k} \, c_{y,j} 
& = &
a^2\Delta_{a}^{k+j}
+ab\Big(\Delta_{a}^{k}\Delta_{b}^{j}
+\Delta_{a}^{j}\Delta_{b}^{k}\Big)
+b^2\Delta_{b}^{k+j}
+\Big(a\Delta_{a}^{k}
+b\Delta_{b}^{k}\Big)\int_{a}^{\bar{b}}\Delta_{s}^{j}\,ds \\[5pt]
& & 
+
\Big(a\Delta_{a}^{j}+b\Delta_{b}^{j}\Big)
\int_{a}^{\bar{b}}\Delta_{s}^{k}\,ds 
+
\left( 
\int_{a}^{\bar{b}}\Delta_{s}^{k}\,ds
\right)
\left(
\int_{a}^{\bar{b}}\Delta_{s}^{j}\,ds
\right), \\[5pt]
\dfrac{
\partial
\left(
c_{y,k} \, c_{y,j}
\right)
}
{\partial a}
& = &
c_{y,j}
\dfrac{\partial c_{y,k}}{\partial a}
+ 
c_{y,k}
\dfrac{\partial c_{y,j}}{\partial a} 
=
a\bar{\Phi}(\gamma)
\dfrac{k \Delta_{a}^{k-1}}{\phi(\Delta_{a})}
c_{y,j}
+ 
a\bar{\Phi}(\gamma)
\dfrac{j \Delta_{a}^{j-1}}{\phi(\Delta_{a})}
c_{y,k} \\[5pt]
& = &
\dfrac{a\bar{\Phi}(\gamma)}{\phi(\Delta_{a})}
\left[ 
k\Delta_{a}^{k-1}c_{y,j}
+ 
j\Delta_{a}^{j-1}c_{y,k}
\right], \\[5pt]
\dfrac{
\partial
\left(
c_{y,k} \, c_{y,j}
\right)
}
{\partial b}
& = &
c_{y,j}
\dfrac{\partial c_{y,k}}{\partial b}
+ 
c_{y,k}
\dfrac{\partial c_{y,j}}{\partial b}
=
-b\bar{\Phi}(\gamma)
\dfrac{k \Delta_{\bar{b}}^{k-1}}{\phi(\Delta_{\bar{b}})}
c_{y,j}
-
b\bar{\Phi}(\gamma)
\dfrac{j \Delta_{\bar{b}}^{j-1}}{\phi(\Delta_{\bar{b}})}
c_{y,k} \\[5pt]
& = &
-\dfrac{b\bar{\Phi}(\gamma)}{\phi(\Delta_{\bar{b}})}
\left[ 
k\Delta_{\bar{b}}^{k-1}c_{y,j}
+ 
j\Delta_{\bar{b}}^{j-1}c_{y,k}
\right].
\end{eqnarray*}

\item 
Similarly, the expressions used in 
Equation \eqref{eqn:SzMWM} are such that
$
\displaystyle 
c_{k}^{*} 
=
\lim_{\gamma \to -\infty} 
c_{y,k}^{*}
$
and are given by:
\begin{align*}
c_{1}^{*} 
& =
c_{2}-c_{1}^2
-a
\dfrac{\partial (c_{2}-c_{1}^2)}
{\partial a}
-b\dfrac{\partial (c_{2}-c_{1}^2)}
{\partial b} 
+a\bar{a}
\left(
\dfrac{\partial c_{1}}{\partial a}
\right)^2
+b\bar{b}
\left(
\dfrac{\partial c_{1}}{\partial b}
\right)^2
-2ab\dfrac{\partial c_{1}}{\partial a}
\dfrac{\partial c_{1}}{\partial b}, \\[5pt]
2 c_{2}^{*}
& =
c_{3}-c_{1}\cdot c_{2}
-a\dfrac{\partial (c_{3}-c_{1}c_{2})}{\partial a}
-b\dfrac{\partial (c_{3}-c_{1}\cdot c_{2})}{\partial b} \\[5pt]
& \qquad\quad 
+a\bar{a}\dfrac{\partial c_{1}}{\partial a}
\dfrac{\partial c_{2}}{\partial a}
+b\bar{b}\dfrac{\partial c_{1}}{\partial b}
\dfrac{\partial c_{2}}{\partial b}
-ab
\left(
\dfrac{\partial c_{1}}{\partial a}
\dfrac{\partial c_{2}}{\partial b}
+\dfrac{\partial c_{1}}{\partial b}
\dfrac{\partial c_{2}}{\partial a}
\right), \\[5pt]
4 c_{3}^{*}
& =
c_{4}-c_{2}^2
-a\dfrac{\partial (c_{4}
-c_{2}^2)}{\partial a}
-b
\dfrac{\partial (c_{4}-c_{2}^2)}{\partial b} 
+a\bar{a}
\left(
\dfrac{\partial c_{2}}{\partial a}
\right)^2
+b\bar{b}
\left(
\dfrac{\partial c_{2}}{\partial b}
\right)^2
-2ab\dfrac{\partial c_{2}}
{\partial a}\dfrac{\partial c_{2}}{\partial b},
\end{align*}
where the involved derivatives are listed below:
\begin{eqnarray*}
\dfrac{\partial c_{k}}{\partial a}
& = & 
\dfrac{
ka 
\left[ 
\Phi^{-1}
(a)
\right]^{k-1}}
{\phi
\left(
\Phi^{-1}
(a)
\right)
}, 
\qquad 
\dfrac{\partial c_{k}}{\partial b}
=
-\dfrac{
kb 
\left[ 
\Phi^{-1}
\left( 
\bar{b}
\right)
\right]^{k-1}}
{\phi
\left(
\Phi^{-1}
\left( 
\bar{b}
\right)
\right)
}, \\[5pt]
\dfrac{
\partial
\left(
c_{k} \, c_{j}
\right)
}
{\partial a}
& = &
\dfrac{a}
{\phi 
\left(
\Phi^{-1}
(a)
\right)
} 
\left[
k 
\left[ 
\Phi^{-1}(a)
\right]^{k-1}
c_{j} 
+ 
j
\left[ 
\Phi^{-1}(a)
\right]^{j-1}
c_{k} 
\right], \\[5pt]
\dfrac{
\partial
\left(
c_{k} \, c_{j}
\right)
}
{\partial b}
& = &
-\dfrac{b}
{\phi 
\left(
\Phi^{-1}
\left(
\bar{b}
\right)
\right)
} 
\left[
k 
\left[ 
\Phi^{-1}
\left(
\bar{b}
\right)
\right]^{k-1}
c_{j} 
+ 
j
\left[ 
\Phi^{-1}
\left(
\bar{b}
\right)
\right]^{j-1}
c_{k} 
\right].
\end{eqnarray*}
\end{enumerate}

\newpage

\section{Simulated Data for Risk Measure Sensitivity Analysis}
\label{sec:AppendixB}

\begin{table}[htb!]
\centering
\begin{tabular}{rrrrrrrrrr}
\hline
70.89 &  290.60 &  47.22  & 100.36 &  411.36 & 1374.10&  1736.60 &   28.12 &  207.47  &  349.15 \\
465.04 &   10.41 &   6.12 &   96.41&   340.27 &  114.82 &  391.35&  1036.80  &  30.69 &   391.35 \\
426.90&   132.63&   64.22 & 2400.00 &  213.66&   257.07&   849.73 &   34.64 &  275.32 &  2257.90 \\
742.44&  1059.80&  556.32&    25.60&  2400.00 &   12.62 &  275.32 & 1288.30 & 2400.00 &    65.85 \\
1014.30 &  232.99&   76.06  & 104.39 &  610.90 &   92.54 & 1403.90 & 1647.10 &  216.80 &   239.70 \\
116.97 & 29.40 & 279.09&  2400.00 &  1718.30&   207.47 &    5.08 &  232.99&     3.02 &  1048.20 \\
132.63&   149.42&  367.44 & 2400.00 &  260.65 &  132.63 &  104.39&   928.74 &   76.06 &   618.03 \\
625.24&   970.69&  54.718 &   45.77 &  670.02 &  198.41  &   3.02&   223.18 & 2400.00&     14.88 \\
81.39&    110.58&  603.83&    33.31 &  358.20 &   34.64&   275.32&     6.12 &   25.60 &   340.27 \\
86.88 &  2098.60&  110.58&  530.62&     53.19&    67.52&    50.18 &   56.26  & 358.20 &    24.36 \\
\hline
\end{tabular}
\caption{Payment $Y$ Sensitivity Analysis Simulated Data.}
\label{tab:Sensitivity_PPY}
\end{table}

\begin{table}[htb!]
\centering
\begin{tabular}{rrrrrrrrrr}
\hline
30.21 & 272.88 &  0  & 336.39 &  99.19 & 284.59&  43.92 &  120.16 &  0 &  445.00\\
6.07&   0 &  45.09 &   41.65&   5.09 &  78.67&  31.11&  1.63  &  0 &   186.15\\
0&   805.80&   17.79 & 569.51 &  0&   38.93&  200.66 &   64.97 &  10.76 &  64.18\\
2.76&  0&  0&    0&  54.65 &   1081.60 & 1479.77 & 0 & 1479.77 &   102.64\\
319.30 & 5.47&   9.79  & 19.42 &  0 &   106.19 & 0 & 458.98 &  33.90 &   0\\
0 & 1479.77 & 0&  361.80&  10.27&   124.25 &  1421.20 &  0& 87.40 &  0\\
90.49&   0&  220.88 & 1479.77 &  92.60 & 6.47 &  0&   50.65 &   15.05 &   67.37\\
2.76&  1479.77&  530.06 &  86.39 &  358.06 &  0  &   1479.77&   29.77 & 13.91&    670.75\\
127.04&    196.41&  50.65&    503.52 &  12.02 &   0&   26.41&  30.21 &  56.04&  8.18\\
124.25 &  0&  08&  0&    0&    218.54&  0 & 145.02  & 48.74&   3.80\\
\hline
\end{tabular}
\caption{Payment $Z$ Sensitivity Analysis Simulated Data.}
\label{tab:Sensitivity_PPZ}
\end{table}

\end{appendices}

\end{document}